\documentclass[prd,nofootinbib,tightenlines,showpacs]{revtex4}
\usepackage[all]{xy}
\usepackage{graphicx}
\def\journal#1, #2, #3#4, #5#6#7#8    {
    {#1~} {#2}  (#5#6#7#8) #3#4}

\begin{document}


\renewcommand{\thesection}{\arabic{section}}
\renewcommand{\theequation}{\arabic{equation}}
\renewcommand {\c}  {\'{c}}
\newcommand {\cc} {\v{c}}
\newcommand {\s}  {\v{s}}
\newcommand {\CC} {\v{C}}
\newcommand {\C}  {\'{C}}
\newcommand {\Z}  {\v{Z}}
\newcommand{\pv}[1]{{-  \hspace {-4.0mm} #1}}

\newcommand{\be}{\begin{equation}} \newcommand{\ee}{\end{equation}}
\newcommand{\bea}{\begin{eqnarray}}\newcommand{\eea}{\end{eqnarray}}
\newcommand{\grad}{\bm \nabla}

\baselineskip=14pt

\begin{flushright}
\hfill{SINP-TNP/06-23}\\
\end{flushright}

\begin{center}
{\bf  \Large Quantization and Conformal Properties of a\\ Generalized Calogero Model}
 
\bigskip

S. Meljanac {\footnote{e-mail: meljanac@irb.hr}}, A. Samsarov {\footnote{e-mail: asamsarov@irb.hr}} \\  
 Rudjer Bo\v{s}kovi\'c Institute, Bijeni\v cka  c.54, HR-10002 Zagreb,
Croatia \\[3mm] 

\bigskip

B. Basu-Mallick{\footnote{email: biru@theory.saha.ernet.in}}, 
Kumar S. Gupta {\footnote{email:kumars.gupta@saha.ac.in}}\\
Theory Division, Saha Institute of Nuclear Physics, 1/AF Bidhannagar, Calcutta 700064, India\\[3mm]

\bigskip

\end{center}
\setcounter{page}{1}
\bigskip


\begin{center}
{\bf   Abstract}

\bigskip

\end{center} 
We analyze a generalization of the quantum Calogero model with the
underlying conformal symmetry, paying special attention to the
two-body model deformation. Owing to the underlying  $ SU(1,1) $
symmetry, we find that the analytic solutions of this model can be
described within the scope of the Bargmann representation analysis and
we investigate its dynamical structure by constructing the
corresponding Fock space realization. The analysis from the standpoint
of supersymmetric quantum mechanics (SUSYQM), when
applied to this problem, reveals that the model is also shape
invariant. For a certain range of the system parameters, the two-body
generalization of the Calogero model
 is shown to admit a one-parameter family of self-adjoint extensions,
 leading to inequivalent quantizations of the system. 
 
\bigskip
\noindent
PACS number(s): 02.30.Ik, 03.65.Fd, 03.65.-w 


\newpage


\section{Introduction}
The structure and application of the Calogero model
 \cite{Calogero:1969xj}, \cite{Olshanetsky:1981dk} and its various
descendants is a subject that receives much attention. Since
these are examples of many-body exactly solvable models which appear 
 in various contexts in physics as well as in
mathematics \cite{poly,qhe,ll,rmt,qet,hs,sw,black,cam,us,us1}, it is of considerable interest 
to find its generalizations which are exactly solvable and integrable. In
particular, it is appealing to investigate whether some sort of
modification of an exactly solvable model will affect its
integrability. 
The dealing with such many-body problems in one dimension has been shown up
as particularly advantageous since there exist several algebraic
techniques that are applicable in this case, due to highly restrictive
spatial degrees of freedom. It is also interesting to find out all possible boundary 
conditions that render the Hamiltonian of the system self-adjoint \cite{reed} and to analyze 
the nature of the corresponding spectrum.

In this paper we investigate one special class of deformation of the
quantum Calogero model. As a prototype, we study the two-body model as
a particularly convenient one for a complete elaboration of the dynamical
structure of the problem, which is basically the same as for all
many-body problems with the underlying $ \;  SU(1,1) \; $ symmetry.
This specific case will be approached from more different directions
that include the Bargmann representation analysis, the ladder
operators formalism and the SUSYQM analysis which is based on the shape-invariance property of the model in question. 
We also study the self-adjoint extensions \cite{reed} of the two-body model and show that
for certain values of the system parameters the model admits a one-parameter
family of inequivalent quantizations.
At the end, the analysis
will be expanded to include the  $ \; N $-body case as well, with the
result for the complete spectrum of the 3-body case stated
explicitly.
The emphasis will particularly be given to the Bargmann representation
approach which
 was put forward in Ref. \cite{Meljanac:2004vi} in context with the
general method for integration of the multi-species and
multi-dimensional generalizations of the Calogero model. This approach
 relies heavily on the conformal invariance of the model
under consideration.


\section{Bargmann representation analysis}
In a recent paper \cite{Meljanac:2004vi}, a general procedure for integrating a many-body
quantum systems with the underlying $ \; SU(1,1) \; $ symmetry was set
up. This procedure is based on the fact that the $ \; N- \; $ body quantum system, possessing
conformal symmetry, can be mapped onto the set of $ \; N \; $ harmonic
oscillators in arbitrary dimensions and with a common frequency $ \;
\omega. $
In this paper we investigate the system described by the
Hamiltonian
\begin{equation} \label{h1}
 H = -\frac{1}{2} \left( \frac{{\partial}^{2}}{\partial x_{2}^{2}} +
 \frac{{\partial}^{2}}{\partial x_{1}^{2}} \right)  + \frac{1}{2} {\omega}^{2}
 (x_{1}^{2} + x_{2}^{2} ) + \frac{\lambda}{2 {(x_{1} - x_{2})}^{2}} +
 \frac{\mu}{2 (x_{1}^{2} + x_{2}^{2} )},
\end{equation}
as a simple example of the model and techniques that were put
forward in \cite{Meljanac:2004vi}. The above Hamiltonian clearly represents two
interacting particles in one-dimensional space bounded by a
harmonic force. For simplicity, masses of the particles are set equal
to $ \; 1 \; $  and $ \; \hbar = 1. $ It is convenient to
  introduce the following set of operators:
\begin{equation}\begin{array}{l} \label{conformalgen}
 T_{+} = \frac{1}{2} \sum_{i=1}^{2} {x_{i}}^{2} =  \frac{1}{2} {r}^{2} , \\ 
 T_{-} = \frac{1}{2}\sum_{i=1}^{2} \frac{{\partial}^{2}}{\partial x_{i}^{2}}
         - V(x_{1}, x_{2} ) = 
    \frac{1}{2} \left( \frac{{\partial}^{2}}{\partial {r}^{2}} 
    + \frac{1}{r} \frac{\partial}{\partial r} + \frac{1}{{r}^{2}}
       \frac{{\partial}^{2}}{\partial {\phi}^{2}} \right) - V, \\
 T_{0} = 
 \frac{1}{2} \sum_{i=1}^{2} {x}_{i} \frac{\partial}{\partial x_{i}}  +
 \frac{1}{2} = \frac{1}{2} r \frac{\partial}{\partial r} + \frac{1}{2},
\end{array}\end{equation}
with  $ \; V \; $ given as
\begin{equation} \label{vpot}
 V = \frac{\lambda}{2 {(x_{1} - x_{2})}^{2}} +
 \frac{\mu}{2 (x_{1}^{2} + x_{2}^{2} )}.
\end{equation}
If we set $ \; \mu = - \lambda  \; $ in (\ref{h1}), we arrive at the
 model considered in Ref. \cite{diaf}.
 The operators $ \;  T_{+},  T_{-} ,  T_{0}\; $ in (\ref{conformalgen})
are expressed in terms of the polar coordinates  $ \; x_{1} = r \sin
\phi, \;\; x_{2} = r \cos \phi. $
A brief inspection of the Hamiltonian (\ref{h1}) reveals that it
belongs to the class of conformally invariant systems \cite{Meljanac:2004vi}, with the
potential $ \; V \; $ being a real homogeneous function of order  $ \; -2, $  
i.e., satisfying the relation
\begin{equation}
  [ \sum_{i=1}^{2} {x}_{i} \frac{\partial}{\partial x_{i}}, V ]  = -2V . 
\end{equation} 
Indeed, it is straightforward to show that the generators (\ref{conformalgen})
 satisfy the SU(1,1) conformal algebra
\begin{equation} \label{conformalalg}
[T_{-},T_{+}] = 2T_{0}, \;\;\;\; [T_{0},T_{\pm}] = \pm T_{\pm}
\end{equation}
and that the Hamiltonian (\ref{h1}) can  be represented in terms of them as
$ \;\; H = -T_{-} + {\omega}^{2}T_{+}. $
The problem of solving for the eigenstates and for the spectrum of the
    Hamiltonian (\ref{h1}), $ \; H \psi_{n,k} = E_{n,k} \psi_{n,k}, \; $ will now be transferred into the
    two-oscillator eigenvalue 
    problem for the Hamiltonian  $ \;\; 2 \omega T_{0}, $ which is usually
    referred to as a transition to a Bargmann representation.
 This can be achieved by applying the transformation                             
\begin{equation} 
 H = 2 \omega  S T_{0} S^{-1},
\end{equation}
where
\begin{equation}  \label{stransformation}
    S = e^{- \omega T_{+}} e^{- \frac{1}{2 \omega} T_{-}}.
\end{equation}
After carrying out the transition to the Bargmann representation, we
 have to solve the eigenvalue
 problem for $ \; T_{0}, $ with an additional constraint 
\begin{equation} \label{toeigenvalue}
  T_{-} \Delta_{n} = 0,  \;\; \;\; T_{0} \Delta_{n} =  \frac{\epsilon_{n}}{2}
\Delta_{n},  \;\; n > 0.
\end{equation}
As it stands, there are many eigenfunctions of  $ \; T_{0}. $ In fact, every
homogeneous function, no matter whether it is rational or irrational,
is an eigenfunction of $ \; T_{0}. $ However, among all of them we have to
pick up only those that are annihilated by $ \; T_{-}. $ These form
an infinite number of
 vacua $ \; \Delta_{n}, \; $ upon which the equidistant towers of
states are built up, with an elementary energy step $ \; 2 \omega \; $
between any two neighbouring states.
The ground state $ \; \psi_{0,0} \; $  of (\ref{h1}) is required to be a square-integrable
function. This will be the case if the ground-state energy 
 $ \; \omega \epsilon_{0}, \;\;  \epsilon_{0} > 0, \;  $
 is higher than $ \;  \frac{\omega}{2}, $ the condition which is connected with  the existence of
the critical point \cite{Meljanac:2003jj, Bardek:2000zj}.
 Then we can write $ \;\; \psi_{0,0} = S \Delta_{0}, \;\; $ where 
$ \Delta_{0} $ is a homogeneous function of the lowest degree and of the lowest energy
 $ \omega  \epsilon_{0}. $ The other vacua $ \; \Delta_{n}, \; n > 0,
 \; $ of higher degrees of homogeneity and with energies  $ \; \omega \epsilon_{n}, \; $ 
are also mapped to  $ \;\; \psi_{n,0} = S \Delta_{n}.\; $
As far as the relations (\ref{conformalalg}) are concerned, the
generators (\ref{conformalgen}) can be viewed as creation and
annihilation operators acting on the eigenstates of $ \; T_{0}. $ In
respect of this,
the excited states $ \; \psi_{n,k} \; $ of (\ref{h1}) can be constructed as  
\begin{equation} \label{strstates}
 \psi_{n,k} = S {T_{+}}^{k}
\Delta_{n},   \;\; k = 0,1,2,...; \;
 n \geq 0, 
\end{equation}
 with energies  $ \;\; 2\omega ( k +  \frac{ \epsilon_{n}}{2} ), 
\; $ and $ \; S \; $ given by (\ref{stransformation}). Thus, to conclude,
 we have all solutions grouped into
equidistant towers of states based on $ S \Delta_{n},
\;\; n \geq 0, $ and for a given
$ n \geq 0, $ the spectrum is equidistant with an elementary step $ 2\omega  $.

It may be noted that the Casimir of the $SU(1,1)$ algebra (\ref{conformalalg}) is given by
\be \label{cas}
{\cal C}  = T_0^2 - T_0 - T_+T_- ,
\ee 
which commutes with the Hamiltonian, the latter being an element of the algebra (\ref{conformalalg}). The given system is thus integrable with the Hamiltonian and the Casimir as the two conserved quantities. By using (\ref{toeigenvalue}), (\ref{strstates}) and (\ref{cas}), we see that
\be \label{cas1}
{\cal C} \psi_{n,k} = \frac{\epsilon_n}{2}\left ( \frac{\epsilon_n}{2} - 1 \right ) \psi_{n,k}.
\ee
Thus it is evident that each tower of states built on the vacuum $\Delta_n$ provides an irreducible representation of the $SU(1,1)$ algebra classified by the eigenvalue of the Casimir ${\cal C}$ given in (\ref{cas1}).

The aforementioned procedure will now be applied to the model Hamiltonian
(\ref{h1}) to yield the eigenstates and the spectrum. First, we have
to solve the equation $ \; T_{-} \Delta_{n} = 0,  \;\; n > 0, \; $ for
the potential (\ref{vpot}),
\begin{equation} \label{annihilationeq}
  \left( \frac{1}{2} \frac{{\partial}^{2}}{\partial x_{1}^{2}} + \frac{1}{2}
 \frac{{\partial}^{2}}{\partial x_{2}^{2}}
  - \frac{\lambda}{2 {(x_{1} - x_{2})}^{2}} -
 \frac{\mu}{2 (x_{1}^{2} + x_{2}^{2} )} \right) \Delta_{n} = 0.
\end{equation}
In polar coordinates this equation reads
\begin{equation} \label{annihilationeq1}
  \left( 
 \frac{ {\partial}^{2} }{\partial r^{2}} +
 \frac{1}{r}  \frac{\partial}{\partial r} + \frac{1}{r^{2}}
   \frac{ {\partial}^{2} }{\partial {\phi}^{2}} -
  \frac{\lambda}{ r^{2}(1 - \sin 2 \phi)} - \frac{\mu}{ r^{2}}
   \right) \Delta_{n} = 0.
\end{equation}
If we follow the lines of Ref. \cite{diaf}, by separating the variables as $ \; \Delta_{n} = u(r) \Phi(\phi), \; $ 
(\ref{annihilationeq1}) reduces to the pair of equations,
\begin{equation}  
 \frac{ {\partial}^{2}u }{\partial r^{2}} +
 \frac{1}{r}  \frac{\partial u}{\partial r} - \frac{C + \mu}{r^{2}}u = 0,
\end{equation}
\begin{equation} \label{kannihilationeq1} 
  \frac{ {\partial}^{2} \Phi }{\partial {\phi}^{2}} -
  \left( \frac{\lambda}{1 - \sin 2 \phi} - C \right) \Phi = 0.
\end{equation}
The first one integrates to $ \; u \sim r^{\sqrt{C + \mu}}, \; $
while the second one appears to have physically acceptable solutions
only for the special values of the coupling $ \; \lambda \; $ and the separation constant $ \; C. $
In order to find these solutions, we skip to the angular variable $ \;
\theta = \phi - \frac{\pi}{4}, \; $ running through the interval $ \;
0 \leq \theta \leq \pi, \; $ and then factorize the function $ \; \Phi, $
\begin{equation} \label{afactorization}
 \Phi_{n} (\theta) = {(\sin \theta )}^{\nu} f_{n} (\cos \theta) = 
                    {(\sin \theta )}^{\nu} f_{n} (x). 
\end{equation}
where the variable $ \; x  = \cos \theta \; $ has been introduced and $ \; n \; $
is the quantum number labeling the vacua $ \; \Delta_{n}. $ 
After inserting (\ref{afactorization}) into equation (\ref{kannihilationeq1}),
we get the equation for the functions $ \; f_{n}, \; n \geq 0  $ 
\begin{equation} \label{eqan1}
   \frac{d^{2} f_{n}(x) }{d x^{2}} {\sin}^{4} \theta
  - ( 2 \nu +1)  x \frac{d f_{n}(x)}{d x} {\sin}^{2} \theta +
   \left( (\nu(\nu - 1) - \frac{\lambda}{2}) x^{2} - (\nu   
    + \frac{\lambda}{2}  - C) {\sin}^{2} \theta \right) f_{n}(x)
  = 0.
\end{equation}
Physically acceptable solutions to the above equation emerge for the
special choices of the coupling and the separation constants $ \;
\lambda \; $ and $ \; C, \; $ respectively,
\begin{equation} \label{physcondition}
 \lambda = 2 \nu (\nu - 1),   \;\;\;\;
 C = {(n + \nu)}^{2},
\end{equation}
in which case Eq. (\ref{eqan1})
     becomes the equation for the Jacobi polynomials $ \;
     P_{n}^{(a,b)} $
\begin{equation} \label{Jacobi}
  (1 - x^{2}) \frac{d^{2} f_{n}(x) }{d x^{2}} 
  - ( 2 \nu +1)  x \frac{d f_{n}(x)}{d x}  +
   n( n + 2 \nu ) f_{n}(x) = 0.
\end{equation}
Thus, as a solution to (\ref{Jacobi}) we have $ \; f_{n}(x) = P_{n}^{(a,a)}(x), \; $ with
$ \; a = \frac{2\nu - 1}{2}. $
Altogether, this gives us the following expression for the vacua in the
 Bargmann representation: 
\begin{equation} \label{Bargmannvacua}
  \Delta_{n} = u \Phi_{n} = r^{\sqrt{C + \mu}} {(\sin \theta )}^{\nu}
  f_{n} (x) = r^{\sqrt{{(n + \nu)}^{2}  + \mu}} {(\sin \theta )}^{\nu}
  P_{n}^{(a,a)}(\cos \theta), 
     \;\;\;\;\;\;  a = \frac{2\nu - 1}{2}.
\end{equation}
In view of eqs. (\ref{conformalalg}), the excited states in the
     Bargmann representation are obtained as
\begin{equation} \label{Bargmannstates}
  {T_{+}}^{k} \Delta_{n} = {(\frac{1}{2} {r}^{2})}^{k}
     r^{\sqrt{{(n + \nu)}^{2}  + \mu}} {(\sin \theta )}^{\nu}
  P_{n}^{(a,a)}(\cos \theta).
\end{equation}
With the  help of the transformation (\ref{stransformation}), we now
     transfer the results back (see eq. (\ref{strstates})) to the original problem to obtain the
     eigenstates of (\ref{h1}),
\begin{equation} \label{nedovrsenatr}
   \psi_{n,k} = S {T_{+}}^{k} \Delta_{n} \sim
      e^{- \omega T_{+}} \sum_{l = 0}^{\infty} \frac{(-1)^{l}}{l!} {(\frac{1}{2\omega} T_{-})}^{l}
     \left( r^{\sqrt{{(n + \nu)}^{2}  + \mu} + 2k} {(\sin \theta )}^{\nu}
  P_{n}^{(a,a)}(\cos \theta) \right).
\end{equation}
For the moment, we shall not border with the proportionality constants
because they are irrelevant for the discussion below.
 This is the reason why the sign of similarity
appears in (\ref{nedovrsenatr}) and in the majority of subsequent
relations.
Later on, when we consider a construction of the bosonic ladder
operators, we shall have to take care of the normalization of the
wave functions and every single proportionality factor will be
important.

 In calculating ( \ref{nedovrsenatr}), we shall make
 use of the fact that the successive application of the
operator $ \; T_{-} \; $ to the states (\ref{Bargmannstates}) leads to an
expression of the form
$$
 {T_{-}}^{l} \left( {T_{+}}^{k} \Delta_{n} \right)
   \sim 2^{-l} 2k \left( 2k + 2\sqrt{{(n + \nu)}^{2} + \mu} \right)
 (2k - 2) \left( 2k - 2 + 2\sqrt{{(n + \nu)}^{2} + \mu} \right) \times
$$
$$
  (2k - 4) \left( 2k - 4 + 2\sqrt{{(n + \nu)}^{2} + \mu} \right) \cdot \cdot \cdot 
  (2k - 2(l - 1)) \left( 2k - 2(l - 1) + 2\sqrt{{(n + \nu)}^{2} + \mu}
  \right) \times
$$
\begin{equation}
 r^{\sqrt{{(n + \nu)}^{2}  + \mu} + 2k - 2l} {(\sin \theta )}^{\nu}
  P_{n}^{(a,a)}(\cos \theta). 
\end{equation}
When $ \; l \; $ is equal to $ \; k, \; $ the above
     expression is, essentially, the Bargmann vacuum $ \; \Delta_{n},
     \; $ up to some constant. Due to the first of Eqs. (\ref{toeigenvalue}), the next application of  $ \; T_{-} \;
     $ yields zero, causing the series in (\ref{nedovrsenatr}) to terminate,
  $ \;  {T_{-}}^{l} \left( {T_{+}}^{k} \Delta_{n} \right) = 0 \; $ for
  $ \; l > k. $
 Relation  ( \ref{nedovrsenatr}) now becomes
$$
 \psi_{n,k} \sim 2^{-k}
  e^{- \frac{\omega}{2} r^{2} } r^{\sqrt{{(n + \nu)}^{2}  + \mu}} {(\sin \theta )}^{\nu}
  P_{n}^{(a,a)}(\cos \theta) \times
$$
$$
 \bigg [ 2^{k} r^{2k} - \frac{2^{k+1}}{2 \omega} k \left(k +
 \sqrt{{(n + \nu)}^{2} + \mu} \right) r^{2k-2} +
 \frac{2^{k+2}}{{(2 \omega)}^{2} 2!}  k(k-1) \left(k +
 \sqrt{{(n + \nu)}^{2} + \mu} \right) \left(k-1 +
 \sqrt{{(n + \nu)}^{2} + \mu} \right) r^{2k-4} + \cdot \cdot \cdot 
$$
$$
  + {(-1)}^{k} \frac{2^{2k}}{{(2 \omega)}^{k} k!}
 k(k-1)(k-2)\cdot \cdot \cdot 2 \cdot 1 
 \left(k +
 \sqrt{{(n + \nu)}^{2} + \mu} \right) \left(k-1 +
 \sqrt{{(n + \nu)}^{2} + \mu} \right) \cdot \cdot \cdot 
  \left(1 +
 \sqrt{{(n + \nu)}^{2} + \mu} \right)  \bigg ]
$$    
\begin{equation} \label{finalequation}
 = { \omega}^{-k} e^{- \frac{\omega}{2} r^{2} }
 r^{\sqrt{{(n + \nu)}^{2}  + \mu}} {(\sin \theta )}^{\nu}
  P_{n}^{(a,a)}(\cos \theta) \sum_{l = 0}^{k}
    {(-1)}^{l} {(\omega r^{2})}^{k-l} {k \choose l} l!
            {k + \sqrt{{(n + \nu)}^{2} + \mu} \choose l},
\end{equation}
where the binomial coefficients in (\ref{finalequation}) have the
following meaning:
$$
 {\alpha \choose \beta} = \frac{\alpha !}{\beta ! (\alpha - \beta)!}
 \equiv \frac{\Gamma(\alpha +1)}{\Gamma(\beta +1) \Gamma(\alpha - \beta + 1) },
$$
with $ \; \Gamma \; $ being the Euler gamma-function.

 If we rearrange the sum in (\ref{finalequation}), we finally obtain the result
$$
 \psi_{n,k} \sim 
  { \omega}^{-k} e^{- \frac{\omega}{2} r^{2} }
  r^{\sqrt{{(n + \nu)}^{2}  + \mu}} {(\sin \theta )}^{\nu}
  P_{n}^{(a,a)}(\cos \theta) \sum_{l = 0}^{k}
    {(-1)}^{k-l} \frac{k!}{l! (k-l)! }
    \frac{ \left( k + \sqrt{{(n + \nu)}^{2} + \mu} \right)! }
  { \left( l + \sqrt{{(n + \nu)}^{2} + \mu} \right)! }
             {(\omega r^{2})}^{l} 
$$
\begin{equation} \label{finalequation1}
 = {(-1)}^{k} k! { \omega}^{-k} e^{- \frac{\omega}{2} r^{2} }
 r^{\sqrt{{(n + \nu)}^{2}  + \mu}} {(\sin \theta )}^{\nu}
  P_{n}^{(a,a)}(\cos \theta)
     L_{k}^{\sqrt{{(n + \nu)}^{2}  + \mu}} (\omega r^{2})
\end{equation}
for the eigenfunctions of the Hamiltonian (\ref{h1}).     
In obtaining this result, the sum in the first line of equation (\ref{finalequation1})
is recognized \cite{abr}, up to the irrelevant factor (anyway, it enters the overall
normalization constant of the wave function), 
 as the power-series expansion for the associated
Laguerre polynomials,
\begin{equation} \label{laguerre}
   L_{k}^{\alpha} (x) = \sum_{m = 0}^{k} \frac{ {(-1)}^{m}}{m!} 
         {k + \alpha \choose  k - m} x^{m};   \;\;\;\;\;  \alpha > -1.
\end{equation}     
Considering the spectrum, as it is unaltered by the transformation
(\ref{stransformation}), we conclude that the eigenenergies of 
(\ref{h1}) are nothing else but the degrees of homogeneity of the
states in the Bargmann representation.
 When $ \; T_{0} \; $ is applied to the Bargmann vacua (\ref{Bargmannvacua}),
we obtain $ \; \frac{1}{2}\left( \sqrt{{(n + \nu)}^{2}  + \mu} + 1
\right) \Delta_{n}  \; $ 
and according to the second one of equations (\ref{toeigenvalue}),
this has to be equal to $ \; \frac{1}{2} \epsilon_{n}. $ Knowing the
dynamical structure of the problem considered ( see Fig. 1), and
anticipating that for a particular vacuum labeled by $ \; n, \; $ the energies
of the neighbouring excited states are separated by $ \; 2 \omega, \; $
the spectrum is easily found to be
\begin{equation} \label{spectrum}
 E_{n,k} = \omega ( \epsilon_{n} + 2k) =  \omega \left(\sqrt{{(n + \nu)}^{2}  + \mu} + 1  + 2k \right),
\end{equation}
with $ \; \nu \; $ being determined by (\ref{physcondition}). 


\section{Ladder operators}
As was said previously, in order to find bosonic ladder operators
\cite{po, borzov, Dadic:2002qn}
related to the eigenfunctions we were in search for, the
knowledge on the detailed form of the normalization factor is essential.
Thus, after we utilize the normalization properties of the Jacobi and
Laguerre polynomials, we are left with the normalized version of the
eigenfunctions (\ref{finalequation1}) of the model Hamiltonian (\ref{h1}),
\begin{equation} \label{finalequationnorm}
 \tilde{\psi}_{n,k} = \sqrt{\frac{ {\omega}^{\sqrt{{(n + \nu)}^{2}  + \mu}+1} k! n! (n + \nu)
      \Gamma (n+ 2\nu)} {4^{\nu - 1} \Gamma \left(\sqrt{{(n + \nu)}^{2}  +
      \mu} + k + 1 \right) {\Gamma (n + \nu + 1/2)}^{2} }}
  e^{- \frac{\omega}{2} r^{2} }
 r^{\sqrt{{(n + \nu)}^{2}  + \mu}} {(\sin \theta )}^{\nu}
  P_{n}^{(a,a)}(\cos \theta)
     L_{k}^{\sqrt{{(n + \nu)}^{2}  + \mu}} (\omega r^{2})
\end{equation}
For brevity, we shall omit the tilde-symbol from the wave
functions  and in the whole
subsequent exposition it will be understood that they  are normalized to unity.
As we are searching for the operators that provide transitions
between ground-state configurations of
 neighbouring towers in Fig. 2, we basically
look for the operators connecting Jacobi polynomials of successive order.
 A straightforward way of deducing on the form of these ladder operators
 (in Fig. 2 they are designated by $ \; B \; $ and $ \; B^{\dagger} $) is to
invoke the recursive relations \cite{abr} for Jacobi polynomials, 
$$
 (2n+a+b)(1- x^{2}) \frac{d}{dx}P_{n}^{(a,b)}(x) =
$$
\begin{equation} \label{jprr}
n[(a-b)-(2n+a+b)x]P_{n}^{(a,b)}(x)+
  2(n+a)(n+b)P_{n-1}^{(a,b)}(x),
\end{equation}
$$
 2(n+1)(n+a+b+1)(2n+a+b)P_{n+1}^{(a,b)}(x) =
$$
\begin{equation} \label{jprr1} 
  (2n+a+b+1)[(2n+a+b)(2n+a+b+2)x+a^{2} - b^{2}]
 P_{n}^{(a,b)}(x)
  -2(n+a)(n+b)(2n+a+b+2)P_{n-1}^{(a,b)}(x).
\end{equation}

Using the recursive relations (\ref{jprr}), (\ref{jprr1}),
we find recursive relations for the normalized ground-state energy eigenfunctions
$ \; \psi_{n,0}, $
$$
 [(n + \nu)x + (1 - x^{2}) \frac{d}{dx} ] \psi_{n,0} =
$$
\begin{equation} \label{recursivedown}
 (n + a) \sqrt{\frac{ {\omega}^{\sqrt{{(n + \nu)}^{2}  + \mu}+1} n! (n + \nu)
      \Gamma (n+ 2\nu)} {4^{\nu - 1} \Gamma \left(\sqrt{{(n + \nu)}^{2}  +
      \mu} + 1 \right) {\Gamma (n + \nu + 1/2) }^{2}  }}
  e^{- \frac{\omega}{2} r^{2} }
 r^{\sqrt{{(n + \nu)}^{2}  + \mu}} {(1 - x^{2})}^{\nu / 2}
  P_{n-1}^{(a,a)}(x),
\end{equation}
$$
 [(n - \nu + 2a + 1)x - (1 - x^{2}) \frac{d}{dx} ] \psi_{n,0} =
$$
\begin{equation} \label{recursiveup}
 = \frac{(n + 1)(n + 2a + 1)}{n + a +1}
 \sqrt{\frac{ {\omega}^{\sqrt{{(n + \nu)}^{2}  + \mu}+1} n! (n + \nu)
      \Gamma (n+ 2\nu)} {4^{\nu - 1} \Gamma \left(\sqrt{{(n + \nu)}^{2}  +
      \mu} + 1 \right) {\Gamma (n + \nu + 1/2) }^{2}  }}
  e^{- \frac{\omega}{2} r^{2} }
 r^{\sqrt{{(n + \nu)}^{2}  + \mu}} {(1 - x^{2})}^{\nu / 2}
  P_{n+1}^{(a,a)}(x),
\end{equation}
where $ \; x = \cos \theta \; $ and $ \; a = \nu - \frac{1}{2}. $

\xy 
\xymatrix{ 
 & n = 0 \ar@{.}[ddd] & & & n = 1 \ar@{.}[ddddd] & & & n = 2
 \ar@{.}[dddd] & & & n = 3 \ar@{.}[ddd] & \ar@{.}[rr] & & \\ \\ \\
 \ar@{-}[rr]^-{|n = 0; k = 3>}   &  & & & & & & & &  \ar@{-}[rr]^-{|n = 3; k = 2>} &  &  \ar@/_3pc/[dlllll]_{B} 
 \ar@/^/[ddd]_{A_{2}^{-}}  \ar@{.}[rr] & &  \\
   & & & & & &  \ar@{-}[rr]^-{|n = 2; k = 2>} & & \ar@/^/[ddd]_{A_{2}^{-}}  \\
   & & & \ar@{-}[rr]^-{|n = 1; k = 2>} & &  \\
   \ar@{-}[rr]^-{|n = 0; k = 2>} \ar@/^/[uuu]_{A_{2}^{+}}
   \ar@/_2pc/[urrrr]_{B^{\dagger}}
  & & & & & & & & &  \ar@{-}[rr]^-{|n = 3; k = 1>}   & & \ar@/^/[ddd]_{A_{2}^{-}}  \ar@{.}[rr] & & \\
   & & & & & &  \ar@{-}[rr]^-{|n = 2; k = 1>}  & &  \\
   & & & \ar@{-}[rr]^-{|n = 1; k = 1>} \ar@/^/[uuu]_{A_{2}^{+}} & &  \\
   \ar@{-}[rr]^-{|n = 0; k = 1>} \ar@/^/[uuu]_{A_{2}^{+}}  & & & & & &
   & & & \ar@{-}[rr]^-{|n = 3; k = 0>} & &
    \ar@/_3pc/[dlllll]_{B} \ar@{.}[rr] & & \\
   & & & & & &  \ar@{-}[rr]^-{|n = 2; k = 0>}
   \ar@/_2pc/[urrrr]_{B^{\dagger}}  & & \ar@/_3pc/[dlllll]_{B} \\
   & & & \ar@{-}[rr]^-{|n = 1; k = 0>} \ar@/_2pc/[urrrr]_{B^{\dagger}}
   & & \ar@/_3pc/[dlllll]_{B}  \\
   \ar@{-}[rr]^-{|n = 0; k = 0>} \ar@/_2pc/[urrrr]_{B^{\dagger}} \ar@/^/[uuu]_{A_{2}^{+}}  & &  }

\endxy

\bigskip

Figure 1: Horizontal shifts in the Fock space of states are accomplished by
the operators $ \; B^{\dagger} \; $ and $ \; B, \; $ whereas the vertical shifts are
provided by the pair of operators  $ \; A_{2}^{+} \; $ and  $ \; A_{2}^{-}. $
\newline
\bigskip
\bigskip

The ladder operators $ \; b \; $ and $ \; b^{\dagger} \; $
 that shift the neighbouring vacua into each other
can be read out from the above recursions
\begin{equation} \label{b}
 b = [x(N + \nu) + (1 - x^{2}) \frac{d}{dx} ]
 \sqrt{\frac{ {\omega}^{\sqrt{{(N + \nu -1)}^{2}  + \mu}}}
   {{\omega}^{\sqrt{{(N + \nu)}^{2}  + \mu}}}
    \frac{ (N + \nu -1) } {(N + \nu) (N + 2\nu -1) }
    \frac{ \Gamma \left(\sqrt{{(N + \nu)}^{2}  +  \mu} + 1 \right)}
     {\Gamma \left(\sqrt{{(N + \nu -1)}^{2}  +  \mu} + 1 \right) } },
\end{equation}
\begin{equation} \label{bdagger}
 b^{\dagger} = [x(N + \nu) - (1 - x^{2}) \frac{d}{dx} ]
 \sqrt{\frac{ {\omega}^{\sqrt{{(N + \nu +1)}^{2}  + \mu}}}
   {{\omega}^{\sqrt{{(N + \nu)}^{2}  + \mu}}}
    \frac{ (N + \nu +1) } {(N + \nu)(N + 2\nu) }
    \frac{ \Gamma \left(\sqrt{{(N + \nu)}^{2}  +  \mu} + 1 \right)}
     {\Gamma \left(\sqrt{{(N + \nu +1)}^{2}  +  \mu} + 1 \right) } },
\end{equation}
with $ \; N \; $ being the number operator defined as
\begin{equation} \label{numberoperator}
 N  \left[ {(1 - x^{2})}^{\nu / 2} P_{n}^{(a,a)}(x) \right]
  = n \left[ {(1 - x^{2})}^{\nu / 2} P_{n}^{(a,a)}(x) \right].
\end{equation}
The explicit form of this operator can be found with the help of Eq.
 (\ref{kannihilationeq1} ) and looks like
\begin{equation} \label{knumber} 
 N = \sqrt{ L^{2} } - \nu;  \;\;\;\;\;\;
   L^{2} \equiv - \frac{ {\partial}^{2}  }{\partial {\theta}^{2}} +
   \frac{\lambda}{1 - \cos 2 \theta}. 
\end{equation}
Since $ \; N \; $ is only angular dependent, relation
(\ref{numberoperator}) can be straightforwardly extended to
\begin{equation} \label{numberoperator1}
 N \psi_{n,0}
  = n \psi_{n,0}.
\end{equation}
Straightforward calculation shows that the ladder operators
$ \; b \; $ and $ \; b^{\dagger} \; $ are bosonic,
\begin{equation} \label{bosonic}
 [b, b^{\dagger}] = 1,
\end{equation}
together with
\begin{equation} \label{bosonicnumber}
 [N, b ] = - b,   \;\;\;\;\; [N, b^{\dagger} ] = b^{\dagger},
\end{equation}
resulting in the simple relation including the number operator $ \; N
= b^{\dagger} b. \; $
However, they are still not the ladder operators we are looking for
because they do not shift the neighbouring eigenstates between each
other, as it is readily seen from (\ref{recursivedown}) and
(\ref{recursiveup}). To overcome this problem, it is
convenient to consider a certain type of operators
that are realized by means of the similarity
transformation applied to the bosonic operators $ \; b \; $ and $ \; b^{\dagger} \; $ 
\begin{equation} \label{transformedb}
 B = r^{\sqrt{L^{2} + \mu}} b r^{- \sqrt{L^{2} + \mu}},
\end{equation}
\begin{equation} \label{transformedbdagger}
 B^{\dagger} = r^{\sqrt{L^{2} + \mu}} b^{\dagger} r^{- \sqrt{L^{2} + \mu}}.
\end{equation}
These operators have the desired properties, namely,
\begin{equation} \label{bosonicrel}
 B \psi_{n,0} = \sqrt{ n} \psi_{n-1,0},   \;\;\;\;\; 
 B^{\dagger} \psi_{n,0} = \sqrt{ n+1} \psi_{n+1,0},
\end{equation}
i.e., they are bosonic, satisfying
\begin{equation} \label{bosonictransformed}
 [B, B^{\dagger}] = 1.
\end{equation}
Owing to the similarity-transformation kind of relation between the operators 
$ \; b,  b^{\dagger} \; $ and $ \; B,  B^{\dagger}, \; $ the latter pair
retains the simple relation to the number operator, namely,
 $ \; N = B^{\dagger} B. \; $
Having established the form of the operators (\ref{transformedb})
and (\ref{transformedbdagger}),
we have succeeded to describe the horizontal shift in the Fock space of states,
depicted in Fig. 1 by horizontal arrows.
The vertical shift in Fig. 1 still remains to be described. However, it is
easily accomplished by the transformed Bargmann-representation
creation and annihilation operators (\ref{conformalgen}),
\begin{equation} \label{verticalladder}
 A_{2}^{\pm} = S T_{\pm} S^{-1} = \frac{1}{2} \left(
 \omega T_{+}  + \frac{1}{\omega} T_{-} \right) \mp T_{0},
\end{equation}
where $ \; T_{+}, T_{-}, T_{0} \; $ are conformal generators (\ref{conformalgen})
and $ \; S \; $ is the transformation (\ref{stransformation}).
Of course, the state of the lowest energy $ \; |0 \rangle \equiv
\psi_{n=0, k=0} \; $ is annihilated by both of the operators 
$ \; B \; $ and $ \;  A_{2}^{-}. \; $
Now, the general Fock-space state of Fig. 1 is easily obtained by the
successive application of the ladder operators
(\ref{transformedbdagger}) and (\ref{verticalladder})
to the vacuum state $ \; |0 \rangle \equiv \psi_{n=0, k=0} $
\begin{equation} \label{fockspace}
 \psi_{n, k} = {(A_{2}^{+})}^{k} { (B^{\dagger})}^{n} |0 \rangle
\end{equation}
and in the coordinate representation it is explicitly described by the
expression (\ref{finalequationnorm}).


\section{SUSYQM analysis of the radial part}
The model under consideration obviously possesses some supersymmetric features
and can be approached from the point of view of the supersymmetric
quantum mechanics  \cite{Cooper:1994eh, Quesne:2003rz, bal}. To see this,
we start with the Hamiltonian (\ref{h1})
\begin{equation} 
 H = -\frac{1}{2} \left( \frac{{\partial}^{2}}{\partial x_{2}^{2}} +
 \frac{{\partial}^{2}}{\partial x_{1}^{2}} \right)  + \frac{1}{2} {\omega}^{2}
 (x_{1}^{2} + x_{2}^{2} ) + \frac{\lambda}{2 {(x_{1} - x_{2})}^{2}} +
 \frac{\mu}{2 (x_{1}^{2} + x_{2}^{2} )}
\end{equation}
By making a transition to polar coordinates and by separating the
variables in the manner as $ \; \Psi(r, \phi) =  u(r) \Phi(\phi), \; $ the Schrodinger equation $
\; H \Psi = E \Psi \; $ becomes
\begin{equation} \label{schmain}
 \frac{1}{u} r^{2} \frac{ {\partial}^{2} u}{\partial r^{2}} +
 \frac{1}{u} r \frac{\partial u}{\partial r} - {\omega}^{2} r^{4} -
 \mu + 2E r^{2} = \frac{\lambda}{1 - \sin 2 \phi} - \frac{1}{\Phi}
 \frac{ {\partial}^{2} \Phi}{\partial {\phi}^{2}} = C,
\end{equation}
where $ \; C \; $ is the separation constant. This separates to the
following two equations:
\begin{equation} \label{req}
 \frac{1}{2} \left( - \frac{ {\partial}^{2} }{\partial r^{2}} -
 \frac{1}{r} \frac{\partial }{\partial r} + {\omega}^{2} r^{2} +
 \frac{C + \mu}{r^{2}} \right) u = Eu,
\end{equation}
\begin{equation} \label{aeq}
 \left( - \frac{ {\partial}^{2} }{\partial {\phi}^{2}} +
 \frac{\lambda}{1 - \sin 2 \phi} -  \right) \Phi  = C \Phi.
\end{equation}
By introducing a new function  $ \; \psi(r) = \sqrt{r} u(r), \; $
(\ref{req}) simplifies to
\begin{equation} \label{req1}
 \frac{1}{2} \left( - \frac{ {\partial}^{2} }{\partial r^{2}}  + {\omega}^{2} r^{2} +
 \frac{C + \mu - \frac{1}{4}}{r^{2}} \right) \psi = E \psi.
\end{equation}
From this equation
 we can immediately read out the radial Hamilton
\begin{equation} \label{rh}
  H_{r} = \frac{1}{2} \left( - \frac{ {\partial}^{2} }{\partial r^{2}}  + {\omega}^{2} r^{2} +
 \frac{C + \mu - \frac{1}{4}}{r^{2}} \right) =  \frac{1}{2} \left( -
 \frac{ {\partial}^{2} }{\partial r^{2}}  + {\omega}^{2} r^{2} +
 \frac{ {\alpha}^{2} - \frac{1}{4}}{r^{2}} \right),
\end{equation}
where the parameter $ \; \alpha = \sqrt{C + \mu} \; $ is introduced,
with $ \; C \; $ determined by (\ref{physcondition}).
The radial Hamiltonian (\ref{rh}) is shape
invariant and this property can be verified by introducing the
corresponding superpotential
\begin{equation} \label{rsuperpot}
 U = \omega r - \frac{\alpha + \frac{1}{2}}{r},
\end{equation}
and by factorizing the Hamiltonian (\ref{rh}) with the help of the
operators $ \; A^{(0)}, \;  {A^{(0)}}^{\dagger}  \; $ introduced in the
following way:
\begin{equation} \label{a}
 A^{(0)} = \frac{1}{\sqrt{2}} \left( \frac{d}{dr} + U \right)
  = \frac{1}{\sqrt{2}} \left( \frac{d}{dr} + \omega r - \frac{\alpha +
 \frac{1}{2}}{r} \right),
\end{equation}
\begin{equation} \label{across}
 {A^{(0)}}^{\dagger} = \frac{1}{\sqrt{2}} \left( - \frac{d}{dr} + U \right)
    = \frac{1}{\sqrt{2}} \left( - \frac{d}{dr} + \omega r -
    \frac{\alpha + \frac{1}{2}}{r} \right).
\end{equation}
Now we have
\begin{equation} \label{rrh}
  H_{r} = {A^{(0)}}^{\dagger} A^{(0)} + e_{0},
\end{equation}
where $ \; e_{0} \; $ is given by (\ref{spectrum}) as $ \; e_{0} = E_{n,0} = \omega(
\alpha + 1). $ 
Let us now create a supersymmetric partner Hamiltonian $ \;H_{r}^{(1)}\; $
 of the Hamiltonian $ \; H_{r} \; \equiv H_{r}^{(0)}, $ 
\begin{equation} \label{rrh1}
 H_{r}^{(1)} = A^{(0)} {A^{(0)}}^{\dagger} + e_{0}.
\end{equation}

\xy 
\xymatrix{ 
  \psi_{n,k}^{(0)}  \ar@{-}[r]
& E_{n,k}^{(0)} \ar@/^2pc/[rr]|-{A^{(0)}}  &
\psi_{n,k-1}^{(1)} \ar@{-}[r] & E_{n,k-1}^{(1)}
\ar@/^2pc/[ll]|-{{A^{(0)}}^{\dagger}} \ar@/^2pc/[rr]|-{A^{(1)}} &
\psi_{n,k-2}^{(2)}
 \ar@{-}[r] & E_{n,k-2}^{(2)} \ar@/^2pc/[ll]|-{{A^{(1)}}^{\dagger}} \ar@{.}[rr]  &
\ar@/^2pc/[rr]|-{A^{(k-1)}}
  & \psi_{n,0}^{(k)} \ar@{-}[r]
& E_{n,0}^{(k)} \ar@/^2pc/[ll]|-{{A^{(k-1)}}^{\dagger}} \\
  \psi_{n,k-1}^{(0)}  \ar@{-}[r]   & E_{n,k-1}^{(0)}
  & \psi_{n,k-2}^{(1)} \ar@{-}[r] &
 E_{n,k-2}^{(1)} \ar@/^2pc/[ll]|-{{A^{(0)}}^{\dagger}} & \psi_{n,k-3}^{(2)}
 \ar@{-}[r] & E_{n,k-3}^{(2)} \ar@/^2pc/[ll]|-{{A^{(1)}}^{\dagger}} \\  \\ \\
 \psi_{n,2}^{(0)} \ar@{-}[r]  & E_{n,2}^{(0)} \ar@{.}[uuu]
\ar@/^2pc/[rr]|-{A^{(0)}}
 & \psi_{n,1}^{(1)}
\ar@{-}[r]  & E_{n,1}^{(1)} \ar@{.}[uuu] \ar@/^2pc/[rr]|-{A^{(1)}}
& \psi_{n,0}^{(2)} \ar@{-}[r]  & E_{n,0}^{(2)} \ar@{.}[uuu] \\
 \psi_{n,1}^{(0)} \ar@{-}[r]  & E_{n,1}^{(0)} \ar@/^2pc/[rr]|-{A^{(0)}} & \psi_{n,0}^{(1)}
\ar@{-}[r]  & E_{n,0}^{(1)} \ar@/^2pc/[ll]|-{{A^{(0)}}^{\dagger}} \\
 \psi_{n,0}^{(0)} \ar@{-}[r]  & E_{n,0}^{(0)} \\
   H_{r} \equiv H^{(0)} & &  H^{(1)} & & H^{(2)} & \ar@{.}[rr] &  
    &  H^{(k)}  }

\endxy

\bigskip

Figure 2: A pattern for constructing the eigenstates of the model
Hamiltonian $ \; H \; $ from the eigenstates of the supersymmetric partner
Hamiltonians $ \; H^{(i)}.$
\newline
\bigskip
\bigskip

The factorization of ( \ref{rrh1}) in a way pursued in (\ref{rrh}) can be achieved by
introducing the operators
\begin{equation} \label{a1}
 A^{(1)} = 
   \frac{1}{\sqrt{2}} \left( \frac{d}{dr} + \omega r - \frac{\alpha +
 \frac{3}{2}}{r} \right),
\end{equation}
\begin{equation} \label{across1}
 {A^{(1)}}^{\dagger} =  \frac{1}{\sqrt{2}} \left( - \frac{d}{dr} + \omega r -
    \frac{\alpha + \frac{3}{2}}{r} \right).
\end{equation}
It is  easy to check that the two sets of operators, (\ref{a}), (\ref{across}) and (\ref{a1}),
(\ref{across1}), satisfy the SUSYQM shape-invariance condition:
\begin{equation} \label{susycon}
 A^{(0)} {A^{(0)}}^{\dagger} = {A^{(1)}}^{\dagger} A^{(1)} +
 e_{1},
\end{equation}
where $ \; e_{1} = 2 \omega. $

A somewhat more detailed insight into the relationships among the
quantities we are dealing with here is obtained by taking a look at Figs. 1
and 2. 
From Fig. 1, let us take an arbitrary tower of states, which, for
the sake of argument, we can take to be
 the tower built up upon the vacuum state labeled by $ \;n. $
This tower coincides identically with the zeroth tower shown in
Fig. 2, with other towers in Fig. 2 being the eigenstates of the
corresponding supersymmetric partner Hamiltonians of the Hamiltonian
(\ref{rrh}). This correspondence is made obvious (see Fig. 2) by labeling each
radial wave function by superscripts in parenthesis. In this way, the
superscript $ \; i  \; $ put on a particular radial function
designates that this function is the eigenfunction of the super
partner Hamiltonian $ \; H_{r}^{(i)}, $ and so on. Each supersymmetric
partner Hamiltonian $ \; H_{r}^{(i)} \; $ has its own set of eigenstates $ \;
\psi_{n,k}^{(i)} \; $ and its own spectrum 
$ \; E_{n,k}^{(i)}, $ with $ \; n, i \; $ fixed and $ \; k \;$  being a
nonnegative integer,
\begin{equation} \label{susyeigeneq}
 H_{r}^{(i)} \psi_{n,k}^{(i)} = E_{n,k}^{(i)} \psi_{n,k}^{(i)},  \;\;\;\;
 k = 0,1,...
\end{equation}
Note that the states from different towers but
in the same horizontal line have the same energy and are
transformed into each other  by
a simple pattern \cite{Cooper:1994eh, Quesne:2003rz}. 

Having the whole picture set up, the basic vacuum ( the state $ \;
\psi_{n,0}^{(0)} $) is obtained from the condition that the operator $
\; A^{(0)}  \; $ should annihilate it,
\begin{equation} \label{basiceq}
  A^{(0)} \psi_{n, k = 0}^{(0)} = 0.
\end{equation}
Equation (\ref{basiceq}) integrates to
\begin{equation}
 \psi_{n,0}^{(0)} = r^{\alpha + \frac{1}{2}} e^{- \frac{\omega}{2}
 r^{2}}.
\end{equation}
In a similar way, the vacuum state of the $ \; i = 1 \; $ tower is
obtained as
\begin{equation}
  A^{(1)} \psi_{n,k = 0}^{(1)} = 0,
\end{equation}
leading to
\begin{equation}
  \psi_{n,k = 0}^{(1)} \sim r^{\alpha + \frac{3}{2}} e^{- \frac{\omega}{2}
 r^{2}}.
\end{equation}
The first excited state $ \; \psi_{n,k = 1}^{(0)} \; $ and the corresponding energy
$ \; E_{n,k = 1}^{(0)} \; $ of the original Hamiltonian (\ref{rh})
follow as
\begin{equation}
 \psi_{n,k = 1}^{(0)} \sim  {A^{(0)}}^{\dagger} \psi_{n,k = 0}^{(1)},
\end{equation}
giving
\begin{equation}
 \psi_{n,k = 1}^{(0)} \sim   e^{- \frac{\omega}{2} r^{2}} 
 r^{\alpha + \frac{1}{2}} [ \omega r^{2} - (\alpha + 1)] =  e^{- \frac{\omega}{2} r^{2}} 
 r^{\alpha + \frac{1}{2}} L_{1}^{\alpha} (\omega r^{2}),
\end{equation}
\begin{equation}
 E_{n,k = 1}^{(0)} = E_{n,k = 0}^{(1)} = e_{0} + e_{1} = e_{0} + 2 \omega,
\end{equation}
where $ \; L_{k}^{\alpha} (\omega r^{2}) \; $ is the associated
Laguerre polynomial and the last equation follows from the combined
application of (\ref{rrh1}), (\ref{susycon}) and (\ref{susyeigeneq}).

By following the same line, we are led (see Fig. 2) to a pattern for constructing a
general radial excitation together with the corresponding excitation
energy. First, we introduce the pair of operators
\begin{equation} \label{ak}
 A^{(k)} = 
   \frac{1}{\sqrt{2}} \left( \frac{d}{dr} + \omega r - \frac{\alpha +
  k + \frac{1}{2}}{r} \right),
\end{equation}
\begin{equation} \label{acrossk}
 {A^{(k)}}^{\dagger} = 
     \frac{1}{\sqrt{2}} \left( - \frac{d}{dr} + \omega r -
    \frac{\alpha + k + \frac{1}{2}}{r} \right).
\end{equation}
The conjecture, now, is that after applying 
(\ref{acrossk}) $ \; j \; $ times successively,
$$
 \psi_{n,j}^{(k-j)} \sim
 {A^{(k-j)}}^{\dagger} \cdot \cdot \cdot  {A^{(k-2)}}^{\dagger}{A^{(k-1)}}^{\dagger}
 \psi_{n,0}^{(k)}, 
$$   
 we get
\begin{equation} \label{generalformula}
 \psi_{n,j}^{(k-j)} \sim r^{\alpha + \frac{1}{2}} e^{- \frac{\omega}{2} r^{2}} 
 r^{k-j} \sum_{s = 0}^{j} {(-1)}^{s} {j \choose s} s! {\alpha+k \choose s} z^{j-s},
\end{equation}
where the variable $ \; z = \omega r^{2} \; $ has been introduced.
This expression can be proved by induction. For  $ \; j = 1, \; $ the
expression (\ref{generalformula}) reduces to 
$$
 \psi_{n,1}^{(k-1)} \sim r^{\alpha + \frac{1}{2}} e^{- \frac{\omega}{2} r^{2}} 
(\omega r^{k+1} - (\alpha + k)r^{k-1}),
$$
which coincides identically with $ \; \psi_{n,1}^{(k-1)} \sim {A^{(k-1)}}^{\dagger}
 \psi_{n,0}^{(k)}. $ For the step of the induction, let us assume that
 (\ref{generalformula}) holds for $ \; j. $ Then we have
$$
 \psi_{n,j+1}^{(k-(j+1))} \sim {A^{(k-j-1)}}^{\dagger} \psi_{n,j}^{(k-j)}
$$
$$
 = \frac{1}{\sqrt{2}} \left( - \frac{d}{dr} + \omega r -
    \frac{\alpha + k - j - \frac{1}{2}}{r} \right)
 \left( r^{\alpha + \frac{1}{2}} e^{- \frac{\omega}{2} r^{2}} r^{k-j} 
 \sum_{s = 0}^{j} {(-1)}^{s} {j \choose s} s! {\alpha+k \choose s} z^{j-s}
 \right)
$$
$$
= \sqrt{2} r^{\alpha + \frac{1}{2}} e^{- \frac{\omega}{2} r^{2}}
r^{k-j-1}
\left(  \sum_{s = 0}^{j} {(-1)}^{s} {j \choose s} s! {\alpha+k \choose s}
  {\omega}^{j-s+1} r^{2j-2s+2} -
  \sum_{s = 0}^{j} {(-1)}^{s} {j \choose s} s! {\alpha+k \choose s} (\alpha + k - s)
  {\omega}^{j-s} r^{2j-2s}
 \right).
$$
By rearranging the sums and by using the properties of the binomial
coefficients, the expression written above becomes
$$
 \sqrt{2} r^{\alpha + \frac{1}{2}} e^{- \frac{\omega}{2} r^{2}}
r^{k-j-1}
 \left( { (\omega r^{2})}^{j+1} + {(-1)}^{j+1} j! {\alpha+k \choose j}
 (\alpha + k - j) +
 \sum_{s = 1}^{j} {(-1)}^{s} {j + 1 \choose s} s! {\alpha+k \choose s}
  { (\omega r^{2})}^{j+1-s}
   \right)
$$
$$
 =  \sqrt{2} r^{\alpha + \frac{1}{2}} e^{- \frac{\omega}{2} r^{2}}
r^{k- (j+1)}
 \sum_{s = 0}^{j+1} {(-1)}^{s} {j+1 \choose s} s! {\alpha+k \choose s}
 { (\omega r^{2})}^{j+1-s},
$$
and this is exactly (\ref{generalformula}) for $ \; j \rightarrow j+1.$
Now that we have verified relation (\ref{generalformula}), the eigenstates of
the radial Hamiltonian $ \; H_{r} \; \equiv H_{r}^{(0)}\; $
follow directly from (\ref{generalformula}) by
setting $ \; j = k, $
\begin{equation} \label{seigenstates}
 \psi_{n,k}^{(0)} \sim r^{\alpha + \frac{1}{2}} e^{- \frac{\omega}{2} r^{2}} 
  \sum_{s = 0}^{k} {(-1)}^{s} {k \choose s} s! {\alpha+k \choose s} z^{k-s}.
\end{equation}
By rearranging the sum in the above expression, (\ref{seigenstates}) is recognized as an
expansion (\ref{laguerre}) for the associated Laguerre polynomials, 
\begin{equation} 
 \psi_{n,k}^{(0)} \sim r^{\alpha + \frac{1}{2}} e^{- \frac{\omega}{2} r^{2}} 
  L_{k}^{\alpha} (\omega r^{2}),
\end{equation}
as it should be, to coincide with the results obtained previously by
pursuing other methods.
To complete the SUSYQM analysis, we turn onto the spectrum of the
model. This analysis, of course, has to yield the same result as
obtained before. 
 While carrying out the consideration, one should note that the shape-invariance condition 
(\ref{susycon}) holds at the general level, i.e.,
\begin{equation} \label{susycongen}
 A^{(k-1)} {A^{(k-1)}}^{\dagger} = {A^{(k)}}^{\dagger} A^{(k)} +
 e_{k}, \;\;\;\; e_{k} = 2 \omega, \;\;\;\; k = 1,2,3,...,
\end{equation}
as well as that the operators $ \; A^{(k)} \; $ annihilate the
corresponding vacua
\begin{equation} \label{annihilation}
  A^{(k)} \psi_{n, 0}^{(k)} = 0, \;\;\;\; k = 1,2,3,...
\end{equation}
Also, as Fig. 2 suggests, the energies of the states in different
towers, but in the same line are the same
\begin{equation} \label{e1e2e3}
 E_{n,k}^{(0)} = E_{n,k-1}^{(1)} = E_{n,k-2}^{(2)} = \cdot \cdot \cdot
 = E_{n,0}^{(k)}.
\end{equation}
Following the construction pattern for the supersymmetric partner Hamiltonians
 $ \;  H_{r}^{(k)}, $ their form is immediately deduced from
the shape-invariance condition (\ref{susycongen})                                     
\begin{equation} \label{rrhk}
  H_{r}^{(k)} = {A^{(k)}}^{\dagger} A^{(k)} + \sum_{j = 0}^{k}
  e_{j},
 \;\;\;\; k = 1,2,3,...
\end{equation}
If we apply (\ref{rrhk}) to $ \; \psi_{n, 0}^{(k)}, $ while
simultaneously anticipating the relations ( \ref{susyeigeneq}),
(\ref{annihilation}) and (\ref{e1e2e3}), we get the spectrum for $ \; H_{r} \; $ as
\begin{equation}
 E_{n,k}^{(0)} = \sum_{j = 0}^{k} e_{j} = e_{0} + 2
 \omega k = \omega ( 2k + \alpha + 1),
\end{equation}
coinciding with (\ref{spectrum}), as expected.

\section{Self-adjoint extension}

In this section we investigate the self-adjoint extensions \cite{reed} of the radial part of the Hamiltonian. Physically this means that we would find all possible boundary conditions for which the radial Hamiltonian is self-adjoint. This would be done by an appropriate analysis of the differential operator for the radial Hamiltonian. We shall see that the system admits a one-parameter family of self-adjoint extensions for certain values of the system parameters. We start with a rederivation of the energy eigenvalues discussed in Section 2 and this method would be carried over when we discuss self-adjoint extensions.

The Schrodinger's equation obeyed by the Hamiltonian (\ref{h1}) of the system is given by 
(\ref{schmain}). The operator involving the radial coordinate $r$ in
the l.h.s. of (\ref{schmain})
 satisfies the eigenvalue equation given by
\be \label{eveqn}
\left [ -\frac{\partial^2}{\partial r^2} - \frac{1}{r} \frac{\partial}{\partial r} 
+ {\omega}^2 r^2 + \frac{\mu + C}{r^2} \right ] u(r) = 2 E u(r),
\ee
where $C = (n+\nu)^2$ as given in (\ref{physcondition}). We shall assume that $n \geq 0$, $\nu \geq 0$.

In order to proceed, we make the following transformations: 
\bea 
u(r) &=& r^\alpha e^{-\frac{\omega r^2}{2}} \chi (r), ~~ \alpha = +\sqrt{\mu + C},\\
t &=& \omega r^2,
\eea
and assume that $\mu + C > 0$.
In these new variables, the radial equation (\ref{eveqn}) becomes
\be \label{trans}
\left [ t \frac{d^2}{d t^2} + (\alpha + 1 - t) \frac{d}{dt} 
- \left (\frac{\alpha + 1}{2} - \frac{E}{2 \omega} \right ) \right ] \chi = 0,
\ee
whose solution is given by \cite{abr}
\be \label{oldsol}
\chi (r) = M \left ( \frac{\alpha + 1}{2} - \frac{E}{2 \omega},~ \alpha + 1,~ \omega r^2 \right ),
\ee
where $M$ denotes the confluent hypergeometric function.
The above solution in general consists of an infinite series. However, in order for the solution to be square integrable, this series must terminate, which happens when
\be \label{termin}
\frac{\alpha + 1}{2} - \frac{E}{2 \omega} = - k,
\ee
where $k$ is a positive integer.
From this, and using the definitions of $b$ and $C$, we obtain that
\be \label{oldevalue}
E_{n,k} = \omega \left(\sqrt{{(n + \nu)}^{2}  + \mu} + 1  + 2k \right),
\ee
which is the same as (\ref{spectrum}). Also, substituting (\ref{termin}) in (\ref{oldsol}) and using the relation between confluent hypergeometric and Laguerre functions \cite{abr}, we see that
\be \label{lag}
\chi (r) = \frac{k!}{(\alpha + 1)_k} L_{k}^{\sqrt{{(n + \nu)}^{2}  + \mu}} (\omega r^{2}),
\ee
where the symbol $ \; (p)_n \; $ means
\be
(p)_n = p (p+1) (p+2) ...... (p+n-1),~(p)_0 = 1.
\ee
The full solution of the radial equation is obtained by substituting (\ref{lag}) in (\ref{trans}). 

We now consider the self-adjoint extensions of the operator $O_r$ where
\be \label{oper}
O_r = \frac{1}{2} \left [ -\frac{\partial^2}{\partial r^2} -
  \frac{1}{r} \frac{\partial}{\partial r} +
 {\omega}^2 r^2 + \frac{\mu + C}{r^2} \right ].
\ee
To begin with, we briefly recall the salient points of von Neumann's theory of self-adjoint extensions \cite{reed}.

        Let $T$ be an unbounded differential operator acting on a Hilbert
space ${\cal H}$ and let $D(T)$ be the domain of $T$. The inner product 
of two elements, $\alpha , \beta \in {\cal H}$ is denoted by $(\alpha ,
\beta)$. Let $D(T^*)$ be the set
of $\varphi \in {\cal H}$ for which there is a unique $\eta \in {\cal H}$ with
$(T \xi , \varphi) = (\xi , \eta )~ \forall~ \xi \in D(T)$. For each such
$\varphi \in D(T^*)$, we define $T^* \varphi = \eta$. The operator $T^*$ then defines the adjoint
of the operator $T$ and $D(T^*)$ is the corresponding domain of the adjoint.
The operator $T$ is called symmetric or Hermitian iff $(T \varphi, \eta) = 
(\varphi, T \eta) ~ \forall ~ \varphi, \eta \in D(T)$. The operator $T$ is called 
self-adjoint iff $T = T^*$ {\it and} $D(T) = D(T^*)$. 

We now state the criterion to determine if a symmetric operator $T$ is
self-adjoint. For this purpose, let us define the deficiency subspaces 
$K_{\pm} \equiv {\rm Ker}(i \mp T^*)$ and the 
deficiency indices $n_{\pm}(T) \equiv
{\rm dim} [K_{\pm}]$. Then $T$ falls in one of the following categories:\\
1) $T$ is (essentially) self-adjoint iff
$( n_+ , n_- ) = (0,0)$.\\
2) $T$ has self-adjoint extensions iff $n_+ = n_-$. There is a one-to-one
correspondence between the self-adjoint extensions of $T$ and the unitary maps
from $K_+$ into $K_-$. \\
3) If $n_+ \neq n_-$, then $T$ has no
self-adjoint extensions.

We now return to the discussion of the operator $O_r$.
This is an unbounded differential operator defined in    
$R^+ $. The operator $O_r$ is a symmetric operator on the domain
$D(O_r) \equiv \{\varphi (0) = \varphi^{\prime} (0) = 0,~
\varphi,~ \varphi^{\prime}~  {\rm absolutely~ continuous},~ \varphi \in {\rm L}^2(rdr)
\} $.
 Next we would like to determine if $O_r$ is self-adjoint. We shall focus on the case where
$\alpha > 0$. 

The deficiency indices $n_{\pm}$ are determined by the number of
square-integrable solutions of the equations
\be \label{defind}
O_r^* u_{\pm}(r) = \pm i u_{\pm}(r),
\ee
respectively, where $O_r^*$ is the adjoint of $O_r$ and the functions
$u_{\pm}(r)$ span the deficiency subspaces $K_{\pm},$ respectively. Note that 
$O_r^*$ is given by the same differential operator as $O_r$.
From dimensional considerations we see that the r.h.s. of Eq. (\ref{defind}) 
should be multiplied by a constant with dimension of ${\rm length}^{-2}$.
We shall henceforth choose the magnitude of this constant to be unity 
by an appropriate choice of units.

Equation (\ref{defind}) can be written as 
\be
\left [ t \frac{d^2}{d t^2} + (\alpha + 1 - t) \frac{d}{dt} 
- \left (\frac{\alpha + 1}{2} \pm \frac{i}{2 \omega} \right ) \right ] \chi_{\pm}(r) = 0,
\ee
where
\be
u_{\pm}(r) = r^\alpha e^{-\frac{\omega r^2}{2}} \chi_{\pm} (r).
\ee
The solutions $u_{\pm}(r)$ of (\ref{defind}) must be square integrable. We are thus led to choose the solutions given by
\be 
\chi_{\pm}(r) = U \left ( g_{\pm}, ~ \alpha+1, ~ \omega r^2 \right ),
\ee
where
\be \label{defsoln}
U \left ( g_{\pm},~ \alpha + 1,~ \omega r^2 \right ) = 
A \bigg [ \frac{M \left ( g_{\pm},~ \alpha+1,~ \omega r^2 \right )}
{\Gamma (d_{\pm}) \Gamma (\alpha+1)}  -  \left ( \omega r^2 \right )^{-\alpha}
\frac{M \left ( d_{\pm},~ 1-\alpha,~ \omega r^2 \right )}
{\Gamma (g_{\pm}) \Gamma (1-\alpha)} \bigg ],
\ee
with $g_{\pm} = \frac{\alpha + 1}{2} \mp \frac{i}{2 \omega}$, $d_{\pm} = 
\frac{1 - \alpha}{2} \mp \frac{i}{2 \omega}$ and $A = 
\frac{\pi}{{\rm sin}(\pi (\alpha +1))}$. Consequently we obtain that
\be
u_{\pm}(r) = r^\alpha e^{-\frac{\omega r^2}{2}}U \left ( g_{\pm},~ \alpha + 1,~ \omega r^2 \right ).
\ee
The functions $u_{\pm}(r)$ are general solutions of (\ref{defind}) with the property that as 
$ r \rightarrow \infty $, $u_{\pm}(r) \rightarrow 0$ sufficiently fast
such that they are square integrable at infinity. In order to
investigate the square integrability of the functions $u_{\pm}(r)$
near $r=0$, first note that as $r \rightarrow 0$, $M(g_{\pm},~b+1,~
\omega r^2) \rightarrow 1$. Using these, we see that as $r \rightarrow 0$,
\be
|u_{\pm}|^2 r dr \label{smallr}
\rightarrow \left [ A_1 r^{(1 + 2 \alpha )} + A_2 r + A_3 r^{(1 - 2 \alpha )} \right ] dr,
\ee
where $A_1, A_2$ and $A_3$ are constants independent of $r$. From (\ref{smallr}) it is clear that the functions $u_{\pm}(r)$ are not square integrable near $r=0$ when $\alpha \geq 1$. Thus when $\alpha \geq 1$, we have the deficiency indices $n_+ = n_- = 0$ and the operator $O_r$ is essentially self-adjoint in the domain $D(O_r)$. However, when $ 0 < \alpha < 1$, the functions $u_{\pm}(r)$ are square integrable near $r=0$ and hence for the whole range of $r$. Thus, when $ 0 < \alpha < 1$, we have the deficiency indices 
$n_+ = n_- = 1$. In this case, according to von Neumann's theory, the operator $O_r$ is not self-adjoint in the domain $D(O_r)$ but admits a one-parameter family of self-adjoint extensions which are labeled by $e^{i z},$ where $z \in R$ (mod $2 \pi$). 
The domain of self-adjointness is given by $D_z(O_r) = D(O_r) \oplus \{ a(u_+ (r) + e^{i z} u_- (r))\},$
 where $a$ is an arbitrary complex number.
 
\begin{figure}
\begin{center}
\includegraphics[width=9cm]{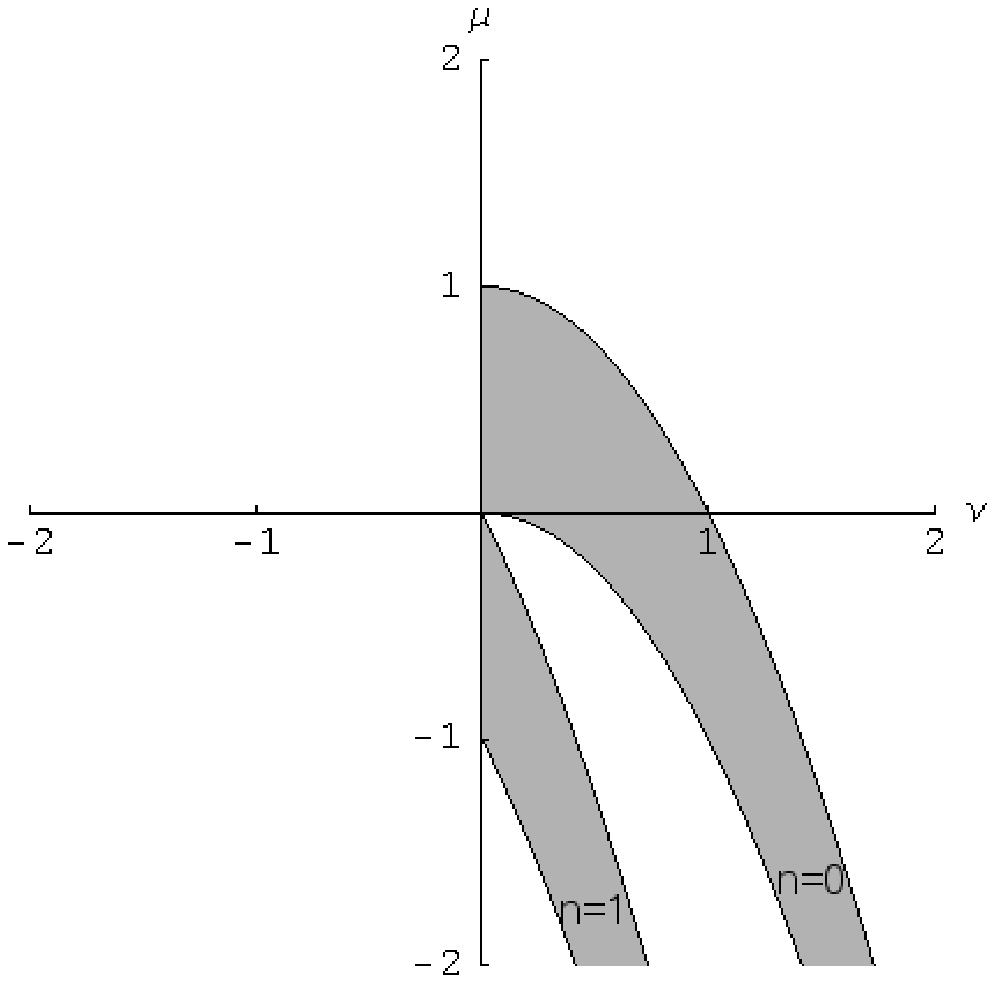}
\end{center}
\begin{flushleft}
Figure 3.
The $n=0$ and $n=1$ bands, representing the 
 parameter ranges where self-adjoint extension is possible, are drawn. 
 Infinite number of such bands exists in the right half of the $\mu -\nu$ plane. 
\end{flushleft}
\end{figure}
Before proceeding, let us discuss the nature of the parameter ranges
for which the system is either essentially self-adjoint or admits 
self-adjoint extensions.  From the condition on $\alpha$, 
we see that the self-adjoint extension exists when
\be \label{saecond}
0 < {\mu + (n + \nu)^2} < 1 . 
\ee
The eigenvalues of the angular equation are characterized by the integer $n \geq 0$. 
For each fixed value of $n$, (\ref{saecond}) gives a region on the right half 
of the $\mu -\nu$ plane for which self-adjoint extensions exist. 
Let us call this region on the $\mu -\nu$ plane as the $n$-th band. 
It is evident that such bands, 
corresponding to different values of $n$, do not overlap with each 
other. Consequently, if the values of the coupling constants $\mu $ and $\nu$
are fixed on the $n$-th band, then the system will admit self-adjoint extension
only if the eigenvalue of the angular equation is taken as $n$. 
It is interesting to note that there exists a finite gap between any two 
consecutive bands labeled by the integers $n$ and $n+1$. Since 
Eq. (\ref{saecond}) is not satisfied for any eigenvalue of the angular equation 
 within such band gaps and also for the region $\mu + \nu^2 >1$, 
the system is essentially self-adjoint in these regions. 
Thus a band structure, with an infinite number of bands
 in the right half of the $\mu -\nu$ plane, represents the parameter
 ranges where the system admits self-adjoint extensions. 
In Fig.3 we have drawn the first two of such infinite number of bands 
 (i.e., $n=0$ and $n=1$ band).

In order to discuss the solutions of the eigenvalue problem, first note that in the parameter range where the system is essentially self-adjoint, the spectrum has already been found before, in  (\ref{oldsol}) and (\ref{oldevalue}).
We now proceed to find the spectrum of $O_r$ in the domain $D_z(O_r)$ for the parameter range where the system admits self-adjoint extensions. To that end, first note that the solution of the eigenvalue equation (\ref{eveqn}) which is square integrable at infinity is given by
\be 
u(r) = B r^\alpha e^{-\frac{\omega r^2}{2}} U \left ( g, ~ \alpha+1, ~
  \omega r^2 \right ),
\ee
where $g = \frac{\alpha + 1}{2} - \frac{E}{2 \omega}$ and $B$ is a constant. In the limit $r \rightarrow 0$, using (\ref{defsoln}), we get that
\be \label{comp1}
u(r) \rightarrow AB \left [ \frac{r^\alpha}{\Gamma (d) \Gamma (\alpha+1)}  
- \frac{{\omega} ^{-\alpha}  r^{- \alpha }}{\Gamma (g) \Gamma (1-\alpha)} \right ],
\ee
where $d = \frac{1 - \alpha}{2} - \frac{E}{2 \omega}$. Again, as $r \rightarrow 0$, we see that
\be \label{comp2}
u_+ (r) + e^{i z} u_- (r) \rightarrow 
A \bigg [ \frac{r^\alpha}{\Gamma (\alpha+1)}  \left ( \frac{1}{\Gamma (d_+) }
+\frac{{\mathrm e}^{iz}}{\Gamma (d_- ) } \right )
  -  \frac{ {\omega} ^{-\alpha} r^{- \alpha }}{\Gamma (1-\alpha)}   
\left ( \frac{1}{\Gamma (g_+) } 
+\frac{{\mathrm e}^{iz}}{\Gamma (g_-) } \right ) \bigg ].
\ee
If $u (r) \in D_z(O_r)$, then by comparing the coefficients of different powers of $r$ in (\ref{comp1}) and (\ref{comp2}) we get,
\be \label{saeen}
f(E) \equiv \frac{\Gamma \left ( \frac{ 1 - \alpha }{2} - \frac{E}{2 \omega} \right )}
{\Gamma \left (\frac{ 1 + \alpha }{2} - \frac{E}{2 \omega} \right ) } \frac{\rho_2 {\mathrm cos}(\frac{z}{2} - \sigma_1)}
{\rho_1 {\mathrm cos}(\frac{z}{2} - \sigma_2)},
\ee 
where
$\Gamma \left ( \frac{ 1 + \alpha }{2} + \frac{i}{2 \omega} \right )
\equiv \rho_1 {\mathrm e}^{i \sigma_1}$
and
$\Gamma \left ( \frac{ 1 - \alpha }{2} + \frac{i}{2 \omega} \right )
\equiv \rho_2 {\mathrm e}^{i \sigma_2}$.
For a given choice of the system parameters, (\ref{saeen}) gives the
energy eigenvalue $E$ as a function of the self-adjoint parameter
$z$. For a fixed set of system parameters, different choices of $z$
lead to inequivalent quantization and to the spectrum for this model in the parameter range where it admits self-adjoint extension. In general, the energy $E$ cannot be calculated analytically and has to be obtained numerically by plotting (\ref{saeen}), a sample of which is given in Fig. 4. 

It is interesting to note that the usual $N$-body Calogero model with the confining interaction leads to a similar radial Hamiltonian, whose self-adjoint extension has been studied before \cite{us1}.  An alternate treatment of the radial problem in the SUSYQM framework, in terms of the self-adjoint extension of the supercharges (first order operators), has been given in \cite{falomir}. Although the radial operators in these works share the same formal structure, the associated physical interpretations vary, e.g. the appearance of the band structure in the parameter space of the present model (as shown in Fig. 3) is absent in the case of the usual Calogero model with confining interaction.

\begin{figure}
\begin{center}
\includegraphics[width=7cm]{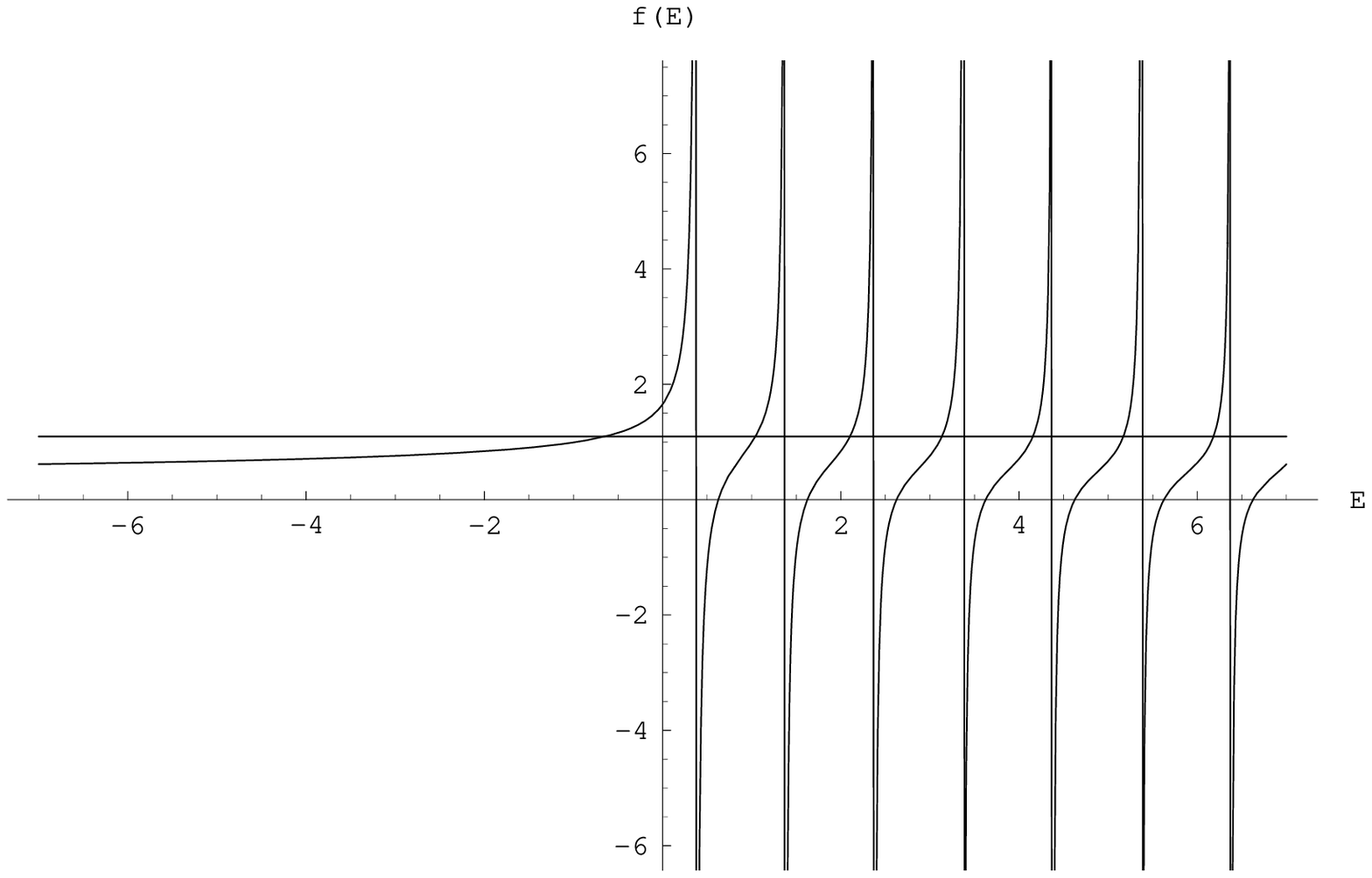}
\end{center}
\noindent
\begin{flushleft}
Figure 4. A plot of Eq. (\ref{saeen}) using Mathematica
with $ \omega = 0.25$, $\alpha = 0.25 $ and 
$z = -1.5$. The horizontal straight line corresponds to the value of the r.h.s of
Eq. (\ref{saeen}).
\end{flushleft}
\end{figure}

We conclude this section with the following observations :
\begin{enumerate}
\item If $\Gamma \left (\frac{ 1 + \alpha }{2} - \frac{E}{2 \omega}
  \right ) = \infty $, we obtain that $E_k =  \omega ( 2 k + \alpha + 1)$ where $k$ is a positive integer. For this to happen, the self-adjoint extension parameter $z$ must take the value $\pi + 2 \sigma_1$ and for this choice of $z$, we recover the usual eigenvalues of this system.
\item For any other choice of $z$, the spectrum of $O_r$ must be obtained numerically and it is seen that the corresponding spectrum is not equispaced in the quantum number $k$. Moreover, generically there is a single negative energy solution. This is due to the fact that for the parameter range where this model admits self-adjoint extension, and for a generic value of $z$, $SU(1,1)$ can no longer be implemented as a spectrum generating algebra.
\item Although we have assumed that $\mu + C > 0$, in general this restriction is not necessary. If this is equal to zero, the system can still be shown to admit a one-parameter family of self-adjoint extensions by using an analysis similar to the one presented here. Also, if $\mu + C$ is strongly negative, then this problem might require renormalization \cite{renorm}, which would be studied elsewhere.
\end{enumerate}


\section{Generalization to the $ \; N \; $ -body case }
 Following the same procedure as for the two-body case, let us now
 investigate some important features of the $ \; N \; $ -body case. In particular,
 we are interested in the general structure of the spectrum
 which basically has the same structure as that shown in Fig. 2,
 except that for the $ \; N \; $ -body case, the Bargmann vacua depend
 on $ \; N-1 \; $ quantum numbers. The problem under consideration is
 described by the Hamiltonian
\begin{equation} \label{Nh}
  H = -\frac{1}{2} \sum_{i = 1}^{N} \frac{{\partial}^{2}}{\partial x_{i}^{2}} +
    + \frac{1}{2} {\omega}^{2}
 \sum_{i = 1}^{N} x_{i}^{2}  + \frac{\lambda}{2} \sum_{i < j} \frac{1}{ {(x_{i} - x_{j})}^{2}} +
 \frac{\mu}{2 \sum_{i = 1}^{N} x_{i}^{2}},
\end{equation}
with $ \; \lambda = 2 \nu(\nu - 1). $
In Ref. \cite{Khare:1995ys} some other deformations of the basic inverse-square
interparticle-distance-model are considered, but these models do not
possess the $ \; SU(1,1) \; $ symmetry. This makes them unsuitable for
the treatment within the scope of the techniques based on the
transition to the
 Bargmann representation.  
 Owing to the underlying
conformal structure of the model (\ref{Nh}), the spectrum and the
eigenstates can be deduced by employing the same
techniques as before. These are based on the fact that the Hamiltonian
(\ref{Nh}) has some special favourable properties which allow us to link it
with the set of decoupled oscillators that is much easier to handle with.
 In order to do this, we introduce the set of
operators
\begin{equation}\begin{array}{l} \label{Nconformalgen}
 T_{+} = \frac{1}{2} \sum_{i=1}^{N} {x_{i}}^{2}, \\ 
 T_{-} = \frac{1}{2}\sum_{i=1}^{N} \frac{{\partial}^{2}}{\partial x_{i}^{2}}
         - \frac{\lambda}{2} \sum_{i < j} \frac{1}{ {(x_{i} - x_{j})}^{2}} -
 \frac{\mu}{2 \sum_{i = 1}^{N} x_{i}^{2}}, \\
 T_{0} = 
 \frac{1}{2} \sum_{i=1}^{N} {x}_{i} \frac{\partial}{\partial x_{i}}  +
 \frac{N}{4}, 
\end{array}\end{equation}
satisfying (\ref{conformalalg}).
The connection of the Hamiltonian (\ref{Nh}) with the set of decoupled
oscillators  \cite{panigrahi}, described by the operator $ \; T_{0}, \; $  is
provided through the transformation (\ref{stransformation}), where now
 $ \; T_{+}, \; $ and $ \; T_{-}, \; $ are from
 (\ref{Nconformalgen}). As a matter of remark, exactly this is what we mean when we are referring to the
 transition to the Bargmann representation. After this transition 
we are left with the equation $ \; T_{-}\Delta = 0, \; $
 which has to be solved in order to find
all Bargmann vacua. Of course, they will now depend on $ \; N-1 \; $
quantum numbers. The remaining quantum number required for the complete
determination of eigenfunctions (eigenfunctions of (\ref{Nh}) are
fully specified with $ \; N \; $ quantum numbers)  will describe
excitations within each tower of states built up over the
corresponding ground state (see Fig. 2). These excitations are governed
by the collective relative radial motion of particles and are a
universal feature \cite{Meljanac:2004vi, Meljanac:2003jj} of all models possessing the underlying $ \; SU(1,1) \; $
symmetry. 

In finding the solution to the equation $ \; T_{-}\Delta = 0, \; $ one can
make a factorization
\begin{equation} \label{factorization}
 \Delta = \prod_{i < j}  {(x_{i} - x_{j})}^{\nu} \phi,
\end{equation}
to obtain an equation for $ \; \phi, $
\begin{equation} \label{Nh1}
  \frac{1}{2} \sum_{i = 1}^{N} \frac{{\partial}^{2}}{\partial
    x_{i}^{2}} \phi +
  \nu \sum_{i < j} \frac{1}{ x_{i} - x_{j}} \left(
    \frac{\partial}{\partial x_{i}} - \frac{\partial}{\partial x_{j}}
    \right) \phi -
 \frac{\mu}{2 \sum_{i = 1}^{N} x_{i}^{2}} \phi = 0.
\end{equation}
One solution to this equation is of the form
\begin{equation} \label{solution0}
  \phi \sim {\left( \sum_{i = 1}^{N} x_{i}^{2} \right)}^{\beta},
\end{equation}
with
\begin{equation} \label{beta}
  \beta = \frac{\nu N - \nu N^{2} - N + 2 + \sqrt{{(N - 2 + \nu N^{2}
  - \nu N )}^{2} + 4\mu}}{4}.
\end{equation}
It is clear that for each ordering of particles the solution
(\ref{solution0}) does not have nodes and hence it is the solution for
the ground state. As far as we are interested in the energy and the
wave function of the
ground state and all vertical (see Fig. 1)
excitations over it, the solution (\ref{solution0}) is all we need to
deduce the spectrum and wavefunctions over the lowest lying vacuum in
Fig. 1. In order to obtain the ground state 
of the model Hamiltonian (\ref{Nh}) and to construct all its
wave functions $ \; \psi_{0,k,0,...,0} \; $ belonging to the first tower of
Fig. 1, we have to use the transformation (\ref{stransformation}) to
return back from the Bargmann representation describing the set of decoupled
oscillators,
\begin{equation} \label{NLaguerre}
 \psi_{0,k,0,...,0} \sim e^{- \omega T_{+}} e^{- \frac{1}{2 \omega} T_{-}}
  \left( {(T_{+})}^{k} \Delta \right)
  = \frac{{(-1)}^{k} k!}{{(2 \omega)}^{k}}
  {\left( \sum_{i = 1}^{N} x_{i}^{2} \right)}^{\beta} \prod_{i < j}  {(x_{i} - x_{j})}^{\nu}
   e^{- \omega T_{+}} L_{k}^{\alpha} (2 \omega T_{+}),
\end{equation}
where $ \; L_{k}^{\alpha} (2 \omega T_{+})  \; $ are the associated
Laguerre polynomials with
\begin{equation} \label{Nspectrum}
 \alpha = \frac{1}{2} \sqrt{{(N - 2 + \nu N^{2} - \nu N )}^{2} + 4\mu}.
\end{equation}
The corresponding
part of the spectrum follows from the equation  $ \; T_{0}\Delta = \frac{\epsilon_{0}}{2} \Delta , \; $
with $ \; T_{0} \; $ and $ \; \Delta  \; $ having the form of
(\ref{Nconformalgen}) and (\ref{factorization}), respectively, and is
given by
\begin{equation} \label{Nspectrum}
 E_{0,k,0,...,0} = \omega (\epsilon_{0,0,...,0} + 2k) =  \omega \left( 1+ \frac{1}{2} \sqrt{{(N - 2 + \nu N^{2}
  - \nu N )}^{2} + 4\mu}  + 2k \right).
\end{equation}
Here
$ \; \epsilon_{0} \; $ is an abbreviation for $ \;
\epsilon_{0,0,...,0} \; $ designating the energy of the ground state,
that is the energy of the lowest of the Bargmann vacua which were seen to
depend on $ \; N-1 \; $ quantum numbers.

Of course,  
 there still remains the problem of finding all solutions to Eq.
( \ref{Nh1}). Obviously, there are many solutions to Eq. ( \ref{Nh1})
and each of them would define one particular Bargmann vacuum.
Once we find all these solutions, we have completed the task of integrating the model
 Hamiltonian (\ref{Nh}) because all excited states 
 built over some particular solution
$ \; \Delta_{n_{1},n_{3},...,n_{N}} \; $ to Eq. ( \ref{Nh1})
 are described by an associated Laguerre
 polynomials in the radial variable $ \; 2 \omega T_{+}. $ So, the
 associated Laguerre polynomial will appear to describe excitations in
 each tower of states shown in Fig. 1
and, as we have seen, this is a common feature for all conformally
 invariant models.

If we look at the problem of integration of  Eq. ( \ref{Nh1})  more
closely, we are naturally led to the question on the
superintegrability \cite{Smirnov} of the Hamiltonian (\ref{Nh}). This question is
closely related \cite{Smirnov} to the problem of separability and even
multi-separability of the corresponding Schrodinger equation. Namely,
a superintegrable system is one that admits more integrals of motion
than it has degrees of freedom, and if it is characterized by a
complete set of commuting quadratic integrals of motion, then it is
also multi-separable. This means that its Schrodinger equation
(i.e. the Hamilton-Jacobi equation in the classical case) allows the
separation of variables in more than one orthogonal system of coordinates.
Since it is known that the  $ \; N \; $ -body Calogero model \cite{Smirnov} is
superintegrable, it would be of interest to see whether the extra term
in (\ref{Nh}) changes anything in this respect.
To get the answer to this question, it is most convenient to consider
 the three-body simplification. If we write the determining condition for Bargmann vacua,
 $ \; T_{-}\Delta = 0, \; $ in spherical coordinates and separate the
 radial part from the angular one, we are confronted with the set
 of relations
\begin{equation} \label{reqN}
   \frac{ {\partial}^{2} }{\partial r^{2}}u +
 \frac{2}{r} \frac{\partial }{\partial r}u  -
 \frac{C + \mu}{r^{2}} u = 0,
\end{equation}
\begin{equation} \label{aeqN}
 \lambda ( G(\theta, \phi) - C )F_{n,m} -  \frac{1}{\sin \theta}
   \frac{ \partial }{\partial \theta} \left( \sin \theta \frac{
   \partial F_{n,m} }{\partial \theta} \right)
 - \frac{1}{{\sin}^{2} \theta} \frac{ {\partial}^{2}
   F_{n,m}}{\partial {\phi}^{2}} = 0,
\end{equation}
where $ \; C \; $ is the integration constant and 
\begin{equation} 
  G(\theta, \phi) = \frac{1}{{\sin}^{2} \theta(1 - \sin 2\phi)}
   + \frac{1}{{(\sin \theta \cos \phi - \cos \theta)}^{2}}  +
 \frac{1}{{(\sin \theta \sin \phi - \cos \theta)}^{2}}. 
\end{equation}
In the above procedure it is understood that the Bargmann vacua
depend on 2 quantum numbers, as is readily expected from the general
consideration, and is separated as
\begin{equation} \label{expressiondelta}
  \Delta_{n,m} = u(r) F_{n,m}(\theta, \phi). 
\end{equation}
While Eq. ( \ref{reqN}) is easily integrated to
$ \; u \sim  \frac{1}{\sqrt{r}} r^{1/2 \sqrt{1 + 4(C + \mu)}}, \; $
the angular equation remains a much more difficult problem to
solve. As has already been said, this problem amounts to the question of
superintegrability of the  3 - body version of the Hamiltonian (\ref{Nh})  and
consequently to the problem of separability of Eq. (\ref{aeqN}). 
  Nevertheless, we can deduce the whole spectrum by relying only on
the expression (\ref{expressiondelta}) and on the fact that the spectrum has to become
a Calogero one in the limit when the parameter  $ \; \mu \; $
approaches zero. With these observations in mind, the whole spectrum
for the 3-body variant of the Hamiltonian (\ref{Nh}) is
readily found to be 
\begin{equation} 
  E_{n,k,m} = \omega \left( 1 + 2k + \frac{1}{2}
 \sqrt{1 + 4 (3 \nu + n + 3m)(1 + 3 \nu + n + 3m ) + 4 \mu} \right). 
\end{equation}


\section{Conclusion}

   We have considered a conformally invariant deformation of the
   quantum Calogero model with the special emphasis on the deformation of the
   two-body model. Owing to the fact that the model under consideration
   possesses $ \; SU(1,1) \; $ symmetry, we were able to apply the
   techniques based on the correspondence between some particular
   conformally invariant model and the set of decoupled oscillators,
   the procedure which is referred to as the Bargmann representation
   analysis. This has
   provided us with the possibility to construct the creation and
   annihilation operators acting on the Fock space, thereby allowing us to investigate the dynamical
   structure of the problem. 

We have also analyzed the self-adjoint extensions of the radial Hamiltonian of the two-body problem and have found the region in the parameter space where the system admits a one-parameter family of self-adjoint extensions. In the situations where the system admits self-adjoint extensions, $SU(1,1)$ can no longer in general be implemented as the spectrum generating algebra and the corresponding spectrum is not equispaced in the quantum number $k$. However, for a special value of the self-adjoint extension parameter, $SU(1,1)$ can be recovered as the spectrum generating algebra, in which case the spectrum becomes equispaced in $k$. It is plausible that the angular part of the Hamiltonian given in (\ref{aeq}) might also admit self-adjoint extensions, anaologous to the angular Hamiltonian of the 3-body Calogero model as discussed in \cite{feher}.

At the end, we have carried out the generalization to include the $ \; N-
   \; $ body problem as well. Here we have found the ground state, which
   is the lowest of all Bargmann vacua, and excitations over it,
   together with the corresponding spectrum. Also,
   we have found that all excitations are basically
   the same, all of them having the origin in the collective relative
   radial motion of the
   particles and are described by the associated Laguerre polynomials in
   the collective radial variable. It appears that this is, in fact, a common
   feature of all models possessing the underlying $ \; SU(1,1) \; $ symmetry.
The only problem, but in no case the simple one, that has been left is to
   find all Bargmann vacua, that is to find all solutions to
   Eq. (\ref{Nh1}). We have seen that this problem is closely related to the
   problem of the superintegrability of the Hamiltonian in question. To
   somehow clear up the situation, we have made the simplification to consider
   the three-body problem. Although we have not found how all Bargmann
   vacua look like in the three-body problem, we were able to deduce
   the complete spectrum for this case. All that was needed was the
   information on the underlying conformal invariance of the model
   and the explicit form of the radial part of the Bargmann vacua wave
   functions, together with the note that the Calogero spectrum has to
   be reproduced in a smooth limit  when the deformation parameter 
   tends to zero.
 We hope that the issues that are left unresolved, as is the
   case with the structure of the Bargmann vacua in a general $ \; N- \; $
   body problem, will be addressed in the near
   future, at least for the three-body case.

{\bf Acknowledgment}\\
B. Basu-Mallick and K. S. Gupta would like to thank N. Bondyopadhaya for assistance with drawing figures. This work was supported by the Ministry of Science and Technology of the Republic of Croatia under 
contract No. 0098003. This work was done within the framework of the
Indo-Croatian Joint Programme of Cooperation in Science and Technology
sponsored by the Department of Science and Technology, India, and
the Ministry of Science, Education and Sports, Republic of Croatia. 



\begin{thebibliography}{99}

\bibitem{Calogero:1969xj}
  F.~Calogero,
  J.\ Math.\ Phys.\  {\bf 10}, 2191(1969);
  F.~Calogero,
  J.\ Math.\ Phys.\  {\bf 10}, 2197(1969);
  F.~Calogero,
  J.\ Math.\ Phys.\  {\bf 12}, 419(1971).
  
  
\bibitem{Olshanetsky:1981dk}
  M.~A.~Olshanetsky and A.~M.~Perelomov,
  Phys.\ Rept.\  {\bf 71}, 313(1981); ibid.{\bf 94}, 313(1983).
  

\bibitem{poly} M. V. N. Murthy and R. Shankar, Phys. Rev. Lett. {\bf 73},
3331 (1994); Z. N. C. Ha, {\it Quantum Many-Body Systems in One
Dimension}, Series on Advances in Statistical Mechanics, Vol. 12, (World
Scientific, 1996);  A. P. Polychronakos, hep-th/9902157; 
B. Basu-Mallick and A. Kundu, Phys. Rev. {\bf B62}, 9927
(2000). 

\bibitem{qhe} H. Azuma and S. Iso, Phys. Lett. {\bf B331}, 107(1994).

\bibitem{ll} N. Kawakami and S.-K. Yang, Phys. Rev. Lett. {\bf 67}, 2493
(1991). 

\bibitem{rmt} B. D. Simons, P. A. Lee, and B. L. Altshuler,
Phys. Rev. Lett. {\bf 72}, 64(1994); S. Jain, Mod. Phys. Lett. {\bf A11},
1201(1996).

\bibitem{qet} C. W. J. Beenakker and B. Rejaei, Phys. Rev. {\bf B49}, 7499
(1994); M. Caselle, Phys. Rev. Lett. {\bf 74}, 2776 (1995).

\bibitem{hs} F. D. M. Haldane, Phys. Rev. Lett. {\bf 60}, 635 (1988); 
B. S. Shastry, Phys. Rev. Lett. {\bf 60}, 639 (1988); A. P. Polychronakos,
Phys. Rev. Lett. {\bf 70}, 2329 (1993).

\bibitem{sw} E. D'Hoker and D. H. Phong, hep-th/9912271; A. Gorsky and
A. Mironov, hep-th/0011197; A. J. Bordner, E. Corrigan and R. Sasaki, Prog.
Theor. Phys. {\bf 100}, 1107 (1998). 

\bibitem{black}
G. W. Gibbons and P. K. Townsend, Phys. Lett. {\bf B454},
187 (1999); D. Birmingham, Kumar S. Gupta and Siddhartha Sen, Phys. Lett. {\bf B505}, 191 (2001); 
Kumar S. Gupta and Siddhartha Sen, Phys. Lett. {\bf B 526}, 121 (2002); Kumar S. Gupta, hep-th/0204137; Kumar S. Gupta and Siddhartha Sen, Phys. Lett. {\bf B 526}, 121 (2002);
Kumar S. Gupta and Siddhartha Sen, Phys. Lett. {\bf B574}, 93 (2003); B. Basu-Mallick, Pijush K. Ghosh and Kumar S. Gupta, Pramana {\bf 62}, 691 (2004).

\bibitem{cam} H. E. Camblong, L. N. Epele, H. Fanchiotti and C. A. Garcia
Canal, Phys. Rev. Lett. {\bf 87}, 220402 (2001).


\bibitem{us}  B. Basu-Mallick and Kumar S. Gupta,
Phys. Lett. {\bf A292}, 36 (2001);  B. Basu-Mallick, Pijush K. Ghosh and Kumar S. Gupta,
Nucl. Phys. {\bf B659}, 437 (2003); Kumar S. Gupta, Mod. Phys. Lett. {\bf A 18}, 2355 (2003).

\bibitem{us1}
 B. Basu-Mallick, Pijush K. Ghosh and Kumar S. Gupta, Phys. Lett. {\bf A311}, 87 (2003).


\bibitem{reed} M. Reed and B. Simon, {\it Methods of Modern Mathematical
Physics}, volume 2, (Academic Press, New York, 1972).


\bibitem{Meljanac:2004vi}
  S.~Meljanac, M.~Milekovic and A.~Samsarov,
  Phys.\ Lett.\ B {\bf 594}, 241 (2004);
  S.~Meljanac and A.~Samsarov,
  Phys.\ Lett.\ B {\bf 613}, 221 (2005)
  [Erratum-ibid.\ B {\bf 620}, 221 (2005)];
  S.~Meljanac and A.~Samsarov,
  Phys.\ Lett.\ A {\bf 351}, 246 (2006).

\bibitem{diaf}A. Diaf, A. T. Kerris, M. Lassaut and R. J. Lombard,
  J.Phys.A:Math.Gen. {\bf  39}, 7305(2006).

\bibitem{Meljanac:2003jj}
S.~Meljanac, M.~Milekovic and A.~Samsarov,                                
Phys.\ Lett.\ B {\bf 573}, 202(2003);
S.~Meljanac, M.~Milekovic, A.~Samsarov and M.~Stojic,
Mod.\ Phys.\ Lett.\ B {\bf 18}, 603(2004);
  S.~Meljanac and A.~Samsarov,
  Phys.\ Lett.\ B {\bf 600}, 179(2004).
 

\bibitem{Bardek:2000zj}
  V.~Bardek and S.~Meljanac,
  Eur.\ Phys.\ J.\ C {\bf 17}, 539(2000);
  S.~Meljanac, M.~Milekovic and M.~Stojic,
  Eur.\ Phys.\ J.\ C {\bf 24}, 331(2002);
 V.~Bardek, L.~Jonke, S.~Meljanac and M.~Milekovic,
 Phys.Lett.B  {\bf  531}, 311(2002). 

\bibitem{abr}
I.~S.~Gradshteyn and I.~M.~Ryzhik,
"Table of Integrals, Series and Products" (Acedemic Press);
M.~Abramowitz and I.~A.~Stegun,
"Handbook of Mathematical Functions, with Formulas, Graphs, and
Mathematical Table" (Dover).
\bibitem{po}
A. P. Polychronakos, Mod. Phys. Lett. A {\bf 5}, 2325(1990);
S.~Meljanac, M.~Milekovi\'c and S.~Pallua,
Phys.\ Lett.\ B {\bf 328}, 55(1994).
\bibitem{borzov}
V. V. Borzov  and E. V. Damaskinsky, "Realization of the annihilation 
operator for generalized oscillator-like system
by a differential operator", math.QA/0101215.

\bibitem{Dadic:2002qn}
  I.~Dadic, L.~Jonke and S.~Meljanac,
   ``Harmonic oscillator with minimal length uncertainty relations and ladder
  operators,''
  Phys.\ Rev.\ D {\bf 67}, 087701(2003).

\bibitem{Cooper:1994eh}
  F.~Cooper, A.~Khare and U.~Sukhatme,
  Phys.\ Rept.\  {\bf 251}, 267(1995);
  P.~K.~Ghosh, A.~Khare and M.~Sivakumar,
  Phys.\ Rev.\ A {\bf 58}, 821(1998);
  C.~J.~Efthimiou and D.~Spector,
  quant-ph/9702017.

\bibitem{Quesne:2003rz}
  C.~Quesne and V.~M.~Tkachuk,
  J.\ Phys.\ A {\bf 36}, 10373(2003);
  C.~Quesne and V.~M.~Tkachuk,
  J.\ Phys.\ A {\bf 37}, 10095(2004).
  
\bibitem{bal}
A.~B.~Balantekin,
Phys.\ Rev.\ A {\bf 57}, 4188(1998);
E.~Drigo Filho and M.~A.~Candido Ribeiro,
``Generalized Ladder Operators for Shape-invariant Potentials,''
nucl-th/0108073.

\bibitem{falomir} 
H. Falomir and P.A.G. Pisani, J. Phys. {\bf A 38}, 4665 (2005).


\bibitem{renorm} K. S. Gupta and S. G. Rajeev, Phys. Rev. {\bf D 48}, 5940 (1993); 
H. E. Camblong, L. N. Epele, H. Fanchiotti, C. A. G. Canal, Phys. Rev. Lett. {\bf 85}, 1590 (2000);
H. E. Camblong, C. R. Ordonez, Phys. Lett. {\bf A 345}, 22 (2005).


\bibitem{Khare:1995ys}
  A.~Khare,
  J.\ Phys.\ A {\bf 29}, L45 (1996).

\bibitem{panigrahi}
N. Gurappa, Prasanta. K. Panigrahi, Phys.Rev. B59  R2490 (1999);
H. Ujino, A. Nishino and M. Wadati, J. Phys. Soc. Jpn. 67 2658 (1998);
P. K. Ghosh, Nucl.Phys. B595 519 (2001);
B.~Basu-Mallick and B.~P.~Mandal,
  Phys.\ Lett.\ A {\bf 284} 231 (2001);
A.~Galajinsky, O.~Lechtenfeld and K.~Polovnikov,
  hep-th/0607215.
  
\bibitem{Smirnov} R.G.Smirnov and P.Winternitz, A class of
  superintegrable systems of Calogero type, math-ph/0606006 and
  references therein.

\bibitem{feher}
L. Feher, I. Tsutsui and T. Fulop, Nucl. Phys. {\bf B 715}, 713 (2005).


\end{thebibliography}
\end{document}